**Fermi National Accelerator Laboratory**

# HUMAN-SYSTEM INTERFACE STYLE GUIDE FOR ACORN DIGITAL CONTROL SYSTEM

Accelerator Controls Operations Research Network

August 2023

Rachael Hill, Madelyn Polzin, Zachary Spielman, Casey Kovesdi, and Dr Katya Le Blanc

ACORN-doc-1475
FERMILAB-TM-2812-AD

## Revision Log

| Revision | Description | Effective Date |
|---|---|---|
| 0.1 | Initial Version, "Style Guide for Accelerator Control Rooms" | 04/21/2023 |
| 0.2 | Human-System Interface Style Guide for ACORN Digital Control System | 08/31/2023 |
| | | |

# Glossary

| Acronym | Term | Definition |
|---|---|---|
| ACNET | Accelerator Control Network | Accelerator Control Network is the primary digital control system utilized at Fermilab. |
| ACORN | Accelerator Control Operations Research Network | The project that will modernize the accelerator control system and replace end-of-life power supplies to enable future operations of the Fermilab Accelerator Complex with megawatt particle beams. |
| BBM | Beam Budget Monitor | Accelerator machinery that measures beam intensity over time. |
| DRS | Design Requirement Specification | A requirement for the system based on meeting design specifications. |
| Fermilab | Fermi National Accelerator Laboratory | US Department of Energy particle physics and accelerator laboratory. |
| HFR | Human Factors Requirements | Requirements that meet the criteria of human factors functional design. |
| HSI | Human-System Interface | The digital interface through which an operator interacts with the accelerator control system. |
| IA | Information Architecture | The overall conceptual model used to plan, structure, and assemble system information. |
| IDA | Information and Decision-Aiding | Logic-based algorithms of automation that assist operators in decision making. |
|  | Comfort Display | Comfort displays are human system interfaces that provide the highest level of information on displays that are viewable by all crew members. |
|  | Page | A specific view or organization of control system information (e.g., index page) |
|  | Parameter | A graphical visualization of accelerator data (usually with an x & y axis) |
|  | Plot | A specific metric/variable of data across all user applications including plots |
|  | Fast Time Plot | A graphical display of parameters (usually 4) of a given machine within a short time frame (0.2 seconds) |
|  | Closure | The end of a beam line |
|  | Tuning | The activity of adjusting magnets and other equipment that interact with a beam of particles to obtain a beam of desired characteristics, e.g., a beam focused onto a target with a small spot size |
|  | Chasing a Loss | The activity of tuning to improve beam quality in one area which often results in higher losses of another as a result |
|  | Ring Wide | Losses around the ring |
|  | Chromaticity | The ratio of tune spread to momentum spread of the beam |
|  | Accelerator Clock | Clocks throughout the accelerator system have their basic frequency directly related to the revolution frequency of the beam (about 7.5 MHz) and produce a clock "tick" every 7 bunches. Beam sync clocks are used for all critical timing of beam transfers between accelerators. |
|  | Parameter Page | A user application that provides custom operations capabilities to consolidate controls and monitoring information into a single location (i.e., a scratch pad) |
|  | Glitch | False beam intensity identified by beam intensity measure up and down stream of glitch |
|  | User Application | An in-house built program |

**Table of Contents**

# Figures

## Tables

# 1.0   SCOPE OF DOCUMENT

## 1.1  Purpose

The purpose of this style guide is to assist developers in designing effective and consistent-looking user interfaces for accelerator control rooms. A similar purpose is to help developers avoid the creation of user interfaces that needlessly stray from the accepted standard set forth in this document. This way, all interfaces combined will look congruous. This is especially beneficial for development that spans multiple years by many different contributors. This document is intended as a ready reference source for all user interface design for the Fermilab accelerator complex.

## 1.2  Scope

This guidance is complementary to ACORN-doc-700, Design Philosophy for Accelerator Control Rooms, as seen in Figure 1.

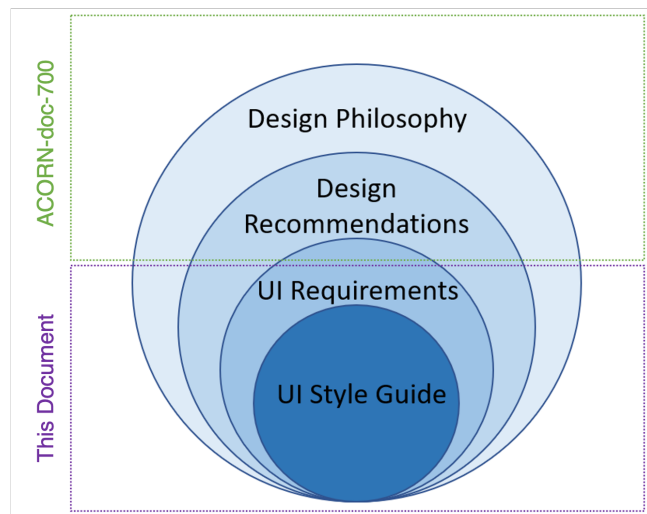

*Figure 1. Process of design philosophy to user interface style guide.*

The guidance in this document, although thorough, is not exhaustive of every design use case a developer might encounter. As such, this style guide is meant to be a "living" document wherein the authors and related collaborators revise and improve upon the established guidance regularly. It is expected that this document will be revised annually at a minimum. As such, areas where specific design guidance is not within scope of the revision are highlighted in gray throughout this document. It is expected that these areas will be completed in later revisions as determined by the project.

## 2.0 PRIMARY SYSTEM USERS

## 2.1 Accelerator Operators

There are multiple types of roles that interact with the accelerator control system including accelerator operators, system experts, and engineering staff. Each of these roles interact with the accelerator control system for specific purposes, and although each role is important, the initial round of interviews focused solely on main control room operators. This means the documented insights included in this report only reflect operator feedback and not machine experts or physicists/engineers. However, it is the intent of the research team to conduct additional interviews to capture insights that represent all roles as a future effort.

### 2.1.1 Crew Chief
This section is out of scope for this revision of the style guide. Future revisions will include display specifications for these roles.

### 2.1.2 Shift Operators
The accelerator shift operator's primary responsibility is to ensure beam viability by acting as a first responder to any accelerator ailments. Should an unexpected event occur, operators act swiftly to mitigate the incident by performing diagnostics and restorative controls. If the needed repair is beyond operator capabilities, machine experts are consulted. Machine experts are specialized to specific equipment and are therefore better equipped to solve complex issues relating to their specific devices (e.g., see Section 2.2.1).

The information a machine expert may be required to diagnose problems with their equipment will differ from the general knowledge held by control room operators. These two roles have similar high-level operational goals (e.g., ensure beam quality), but they address those goals with access to different levels of information. A summary of key functions performed by operators is presented in the task analysis results (Section 7.0).

### 2.1.3 Operations Specialist Staff
This section is out of scope for this revision of the style guide. Future revisions will include display specifications for these roles.

## 2.2 System Experts

### 2.2.1 Machine Experts
This section is out of scope for this revision of the style guide. Future revisions will include display specifications for these roles.

### 2.2.2 Physicists

This section is out of scope for this revision of the style guide. Future revisions will include display specifications for these roles.

## 2.3 Engineering Staff

### 2.3.1 Integrators

This section is out of scope for this revision of the style guide. Future revisions will include display specifications for these roles.

### 2.3.2 Engineers

This section is out of scope for this revision of the style guide. Future revisions will include display specifications for these roles.

### 2.3.3 Technicians

This section is out of scope for this revision of the style guide. Future revisions will include display specifications for these roles.

# 3.0 KEY REFERENCES

# 4.0 HUMAN FACTORS GLOBAL STYLE GUIDE

This section describes global human factors requirements (HFRs) that apply to the hardware and software for the control system. These HFRs are described within the context of 1) information and navigation, 2) display formatting, and 3) controls and system interaction.

## 4.1 Information Architecture and Navigation

An accelerator complex includes a variety of information from multiple sources. To accommodate the complexity and variety of information included in an accelerator complex as well as present that information in ways that are intrinsically meaningful to operators, a well-planned structure is needed. The following sections provide general recommendations and details for how to organize the information included in accelerator control system interfaces. The purpose of an information architecture is to organize and construct system information in a way that simplifies user comprehension to optimize operational performance. A successful information architecture helps users to understand where they are in a system, what they're interacting with, how to navigate their interaction, and what to expect next.

### 4.1.1 Information Architecture

**[HFR-1]** **A two-level hierarchy should be used for system display organization.**
*Details:* In a particle accelerator control system environment, two levels support an intuitive and hierarchically organized information architecture (IA): comfort displays (overview, high-level/ information only, i.e., Level 1) and operator interfaces (lower-level information and operator control; Level 2). This is illustrated in Figure 2. *Technical Basis:* 0 (Section 4.1)

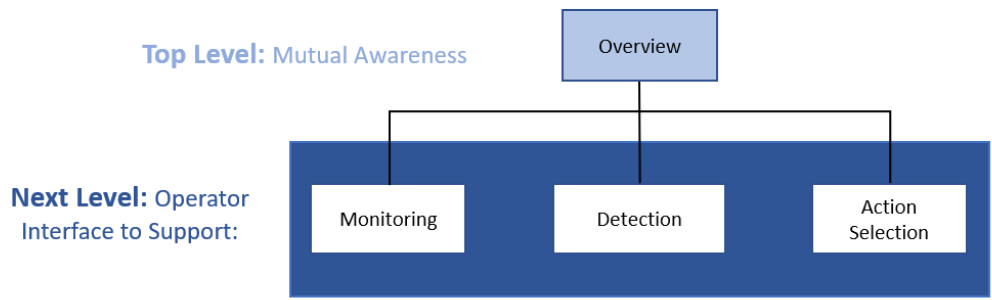

*Figure 2: Accelerator tree structure (adapted from 0). Top Level: Mutual awareness [HFR-2] [HFR-3] Next Level: Operator selected/pertinent/consistent information [HFR-4] [HFR-5].*

**[HFR-2]** **Level 1 information should provide general accelerator information that supports situation awareness of equipment status and beam health.**

*Details:* Operators currently use "Comfort Displays" to provide overview information. These displays provide key parameters important for monitoring. *Technical Basis:* 0 (Section 4.1)

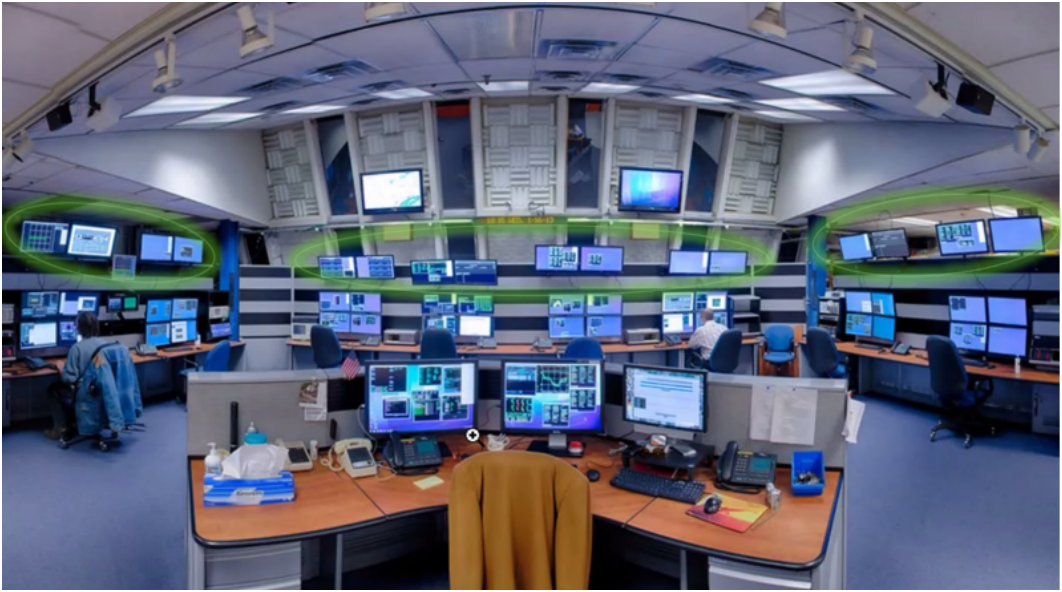

*Figure 3. Accelerator overview displays highlighted in green ("comfort displays").*

**[HFR-3]** **Level 1 information should be presented on the 'Comfort Display' to rapidly communicate degrading, abnormal, or emergency conditions from "at-a-glance."**

*Details: Comfort displays* are the top level of accelerator information in the proposed information hierarchy. The comfort displays support broad system awareness by abstracting and aggregating accelerator information for at-a-glance status acquisition. Operators should rely on comfort displays to provide the first indications of degrading, abnormal, or emergency conditions for accelerator monitoring*. Technical Basis:* 0 (Section 4.2)

**[HFR-4]** **Level 2 information should support the operator in performing their tasks to accomplish the functions assigned to them.**

*Details:* The level 2 accelerator interfaces should support a variety of operator tasks. Each interface should be organized contextually according to user responsibilities and the appropriate level of information and functionality to accommodate said responsibilities. *Technical Basis:* 0 (Section 4.3)

*Note:* This HFR is met by addressing the requirements and design requirement specifications presented in Section 5.0 (i.e., that are a result of function/ task analysis), provided for each accelerator key function.

### 4.1.2 Navigation

Navigation is the method used to locate information within an interface. Navigation is also how a user can get from point A to point B in the most intuitive and efficient way possible. Properly designed navigation not only provides visual indications for where a user can go, but also for where the user is at currently. Navigation design encompasses both visual and functional design.

**[HFR-5]**      **The system should provide an index page that is organized by 1) function and 2) frequency of use for alternative navigation.**

*Details:* A current pain point is that the programs of index pages are organized alphabetically instead of being prioritized in place of function or frequency of use. *Technical Basis:* Key Insights (Insight 5A)

**[HFR-6]**      **System navigational options should be visible on all pages.**

*Details:* Each page should have visible navigational options. Navigational options for pages include forward (i.e., next page), backward (i.e., previous page), minimize (i.e., reduce visibility without closing out), restore up/down (i.e., make window bigger or smaller), and exit (i.e., close out). *Technical Basis:* Key Insights (Insight 5A)

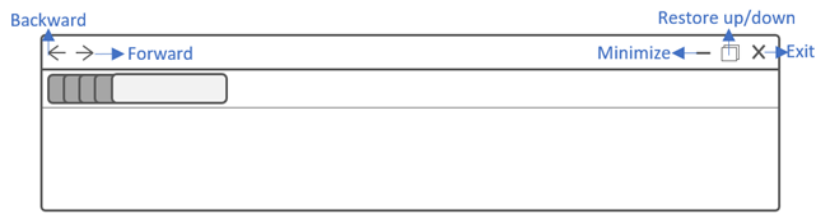

*Figure 4. Navigational options design.*

**[HFR-7]**      **The system should provide visual cues to inform user of where they are in the system.**

*Details:* Users should always know where they are in a digital system. There are multiple ways to indicate current location to a user. A common approach to this is to design a traceable path (e.g., tabs, pagination, and breadcrumbs). *Technical Basis:* Key Insights (Insight 5A)

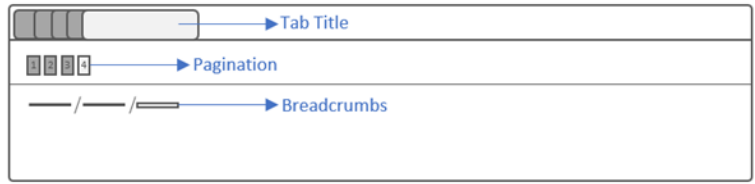

*Figure 5. Common examples of visual cues to indicate current user location.*

**[HFR-8]**    **The system should clearly differentiate navigational elements from each other but visually group similar navigational elements together.**

*Details:* Different types of navigational elements should look different from each other but should be consistently accessible from an expected location on the interface. Group like navigational elements together (e.g., forward/backward and exit/minimize) to increase the consistency and predictability of an interface. . *Technical Basis:* Key Insights (Insight 5A)

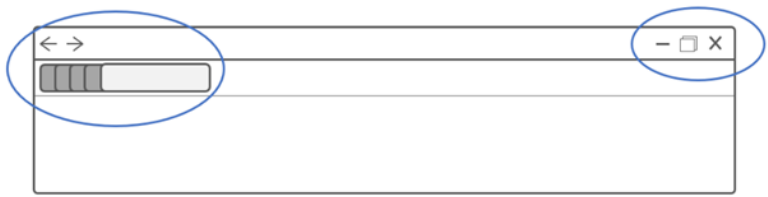

*Figure 6. Grouping similar navigational elements together.*

### 4.1.3  Windows, Pop-Ups, Pages, and Faceplates

Windows, pop-ups, and faceplates are used somewhat synonymously throughout web application literature but are described with distinct definitions that have been created for the purposes of this document. Windows are two dimensional rectangles that can be resized, moved, hidden, restored, or closed. Windows can overlap the visual area of other windows (i.e., occlude each other). 0

**[HFR-9]**    **The content provided on all windows, pop-ups, and faceplates of the system should be located in a consistent location.**

*Details:* The content of displays and their formats should be consistent within a system and across systems that are used by the same users. This includes a consistent header, layout, means for closing windows, entering data, and navigating. *Technical Basis:*  0 (Section 5.17.15.3.1)

**[HFR-10]**   **Pop-ups and faceplates should be presented in a location that does not obscure information related to operating equipment.**

*Details:* Wherever the pop-up (e.g., control faceplate) is placed, it must not block visibility to the information used to monitor the resulting outcome information. If possible, also consider if important downstream information is blocked by the pop-up. *Technical Basis:* 0 (Section 6.4.1.2)

### 4.1.4 Menus And Search

A navigation menu is a section of an interface (i.e., typically a horizontal or vertical bar in the top of the interface) that increases the efficiency and accessibility of navigation throughout a digital system (Farrell, S. 2015). Menus can be designed as fixed menus, drop-down menus, and search menus. Each menu type assists in navigational capabilities of an interface. Fixed menus always displayed (i.e., not collapsible) and are appropriate to use when each navigational option must be accessed frequently or quickly. Drop down menus are appropriate to use when developers are trying to conserve screen space but still provide quick access to navigational options (i.e., a click away). Search menus are appropriate to use when a user must access specific system information that is not readily available. The figures below include design examples of a fixed menu, drop-down menu, and search menu.

**[HFR-11]** **The primary display page should provide a search capability.**

*Details:* The troubleshooting and problem solving that occurs within accelerator operations is highly variable. As a result, having a diverse means to navigate the control system to access information and controls is important. *Technical Basis:* Key Insights (Insight 9A)

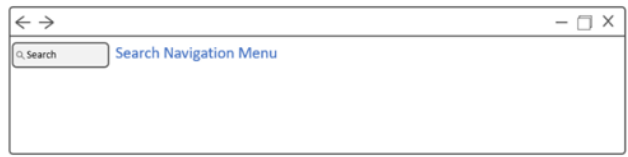

*Figure 7. Search navigation menu (top-left).*

**[HFR-12]** **The system should provide a drop-down menu across all display pages.**

*Details:* The accelerator interface should include an accessible and clearly labeled menu to optimize user comprehension and navigation effectiveness. *Technical Basis:* 0 (Section 6.1.2)

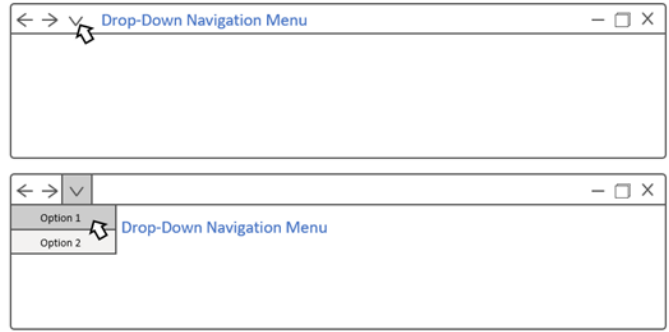

*Figure 8. Drop-down navigation menu (top-left).*

**[HFR-13]** **The drop-down menu should be located in the top-left corner of the header across all display pages.**

*Details:* When the web was created and a broader population of people had access to a digital interaction platform, navigation menus were developed to conserve space on the screen. Due to conformity and convention of top-to-bottom and left-to-right, the top bar of a screen became a standard location for navigation menus. To avoid straying from common web conventions, menus should be located in the top bar of the current page. Most digital interfaces designate the top-left corner for menu placement. When menus are placed outside of these areas, it can feel awkward and counterintuitive from a usability standpoint. *Technical Basis:* 0

**[HFR-14]** **A drop-down menu should contain category labels from which the drop-down menus is accessed.**

*Details:* An effective menu is a menu that includes labels that are intrinsically meaningful and plainly described. Menus are only as clear as the labels and contextual terminology included in them. Labels included in a menu should follow the same guidance recommended in the labels section: plain and concise yet descriptive language that is intrinsically meaningful. *Technical Basis:* 0 (Section 6.1.2); 0 (Section 5.17.3.1.1)

## 4.2 Display Formatting

### 4.2.1 Display Layout

The layout of an HSI is the arrangement of visual elements to achieve a specific goal or goals. The layout of an HSI is commonly determined by prioritization of interface content. How an interface is organized determines user comprehension. A well-structured layout increases a user's ability to effectively understand and navigate an interface.

**[HFR-15]** **All display pages should contain a header with a unique title at the top of the page.**

*Details:* Each display shall be labeled with a title or header. The title should be unique within the system and positioned at the top (left corner or centered). Titles should be intrinsically meaningful with plain language that is concise yet descriptive to optimize user comprehension and interaction effectiveness. Labels should reflect the interface information and components associated to the label. *Technical Basis:* 0 (Section 5.17.15.6; Sections 5.17.15.6.1 – 5.17.15.6.4); 0 (Section 1.5-3)

**[HFR-16]** **All display pages should provide a navigation menu at the top left within the header.**

*Details:* A navigation menu is a section of an interface that increases the efficiency and accessibility of navigation throughout a digital system. To avoid straying from common web conventions, menus should be located in the top bar of the current page. Different types of navigational elements should look different from each other but should be consistently accessible from an expected location on the interface. Group like navigational elements together (e.g., forward/backward and exit/minimize) to increase the consistency and predictability of an interface. *Technical Basis:* 0 (Section 6.1.2) 0

**[HFR-17]** **All display pages should provide a selectable breadcrumb.**

*Details:* Users should always know where they are in a digital system. A selectable breadcrumb acts as another source of navigation for the user with a traceable path. *Technical Basis:* 0 (Section 6.2.2.1)

**[HFR-18]** **All display pages should provide primary canvas area that is consistently sized to support the user's primary task.**

*Details:* Display pages need to be the appropriate size for the task at hand. A page's layout will depend on the content and will be displayed to best support the user. Having too much on one page in a small canvas area can limit the size of the content sections and reduce visibility. Too much unused white space takes up room on a screen that could go to something else and could interfere with how one sets up their screen. *Technical Basis:* 0 (Section 1.5-1)

**[HFR-19]** **All display pages should provide a unique display number that can be referenced through search entry.**

*Details:* A unique display alphanumeric code is one way to identify a page. This shorthand version allows users to navigate to the page they need via search and obtain faster results. Some titles of pages may be similar based on content and a unique display code helps differentiate the pages. This can help in the way of multiple machines having applications with similar functionality. *Technical Basis:* 0 (Section 1.5-4)

**Display Page Customization**

**[HFR-20]** **The system should provide the capability of saving custom display configurations (e.g., multiple graphs).**

*Details:* Due to the complexity of the control system, users monitor different plots based on their workflow and user needs. The ability of saving custom display configurations supports their workflow and tasks. A user's display setup can save time and reduce the number of actions they need to take to access what is regularly needed. *Technical Basis:* Key Insights (Insight 5B)

**[HFR-21]** **Custom display configurations should be tied to a user's login.**

*Details:* Most pages included in a digital control system are fixed (i.e., the content included within is always the same) however some are flexible (i.e., content varies according to user selection). For flexible pages, there should be a list of contents that users can select to populate the page (hence the variability between user visits). *Technical Basis:* Key Insights (Insight 5B)

**[HFR-22]** **The means of saving a display configuration should be explicitly visible to the user.**

*Details:* On pages in which users can save content, the save feature should be visible and consistently located in the same place on all savable pages. Through an autosave feature or a button the user clicks, the act of saving should be conveyed to offer reassurance and trust from the user. *Technical Basis:* Key Insights (Insight 5B)

### 4.2.2 Color

In all types of interfaces, colors are used to indicate meaning, draw a user's attention, and visually separate parts of an interface. In environments as complex as an accelerator control system, it might seem logical to use aa different color for every different field, object, or piece of equipment or data to distinguish their meaning. However, the more saturated colors included in an interface, the higher likelihood the same colors can be reused for different meanings, sometimes of which are conflicting. An overuse of color can also lead to disguising all types of information included in an interface as equally important. For example, alarms can go undetected on an interface already full of highly saturated colors, especially if the alarm color is used for something else on the interface. The amount of colors appropriate for a digital display according to International Society of Automation Standards (ISA 5.5) state that the number of colors should be limited to the minimum required for the display's objective. In other words, irrelevant color use should be avoided at all costs to reduce the visual saturation of a display which in turn alleviates user cognitive burden. Avoiding overuse or irrelevant use of color also strengthens the color-coding associations of an interface. Specific guidelines concerning color use for accelerator interfaces are included below.  0

**[HFR-23]** **Color should be used appropriately to indicate meaning.**

*Details:* Each color included in an accelerator interface should only have one meaning consistently throughout the system. Multiple meanings used for a single-color can cause confusion and performance issues. When multiple colors are used in a single interface, users must remember each of the different meanings in addition to their work tasks. The reliance on a user's memory is even further complicated when a single color is used to convey multiple meanings. To improve the comprehensibility of color-coding associations in accelerator interfaces, each color must be linked to a single meaning or purpose. *Technical Basis:* 0 (Section 6.3.1); 0 (Sections 7.2.5.1 – 7.2.5.10); 0 (Section 1.3.8)

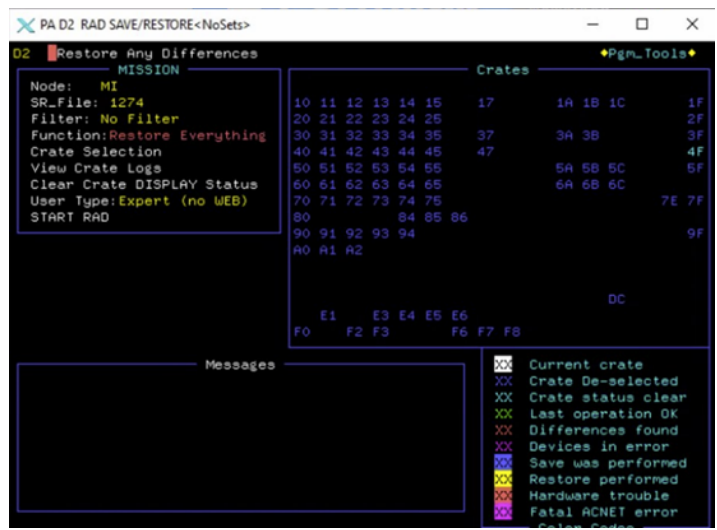

*Figure 9. Poor example of color use. One color (I.e., teal blue) is used to indicate multiple meanings (I.e., "crate status clear" and general text/headings). Each color should only have one meaning and general text colors shouldn't be reused to indicate a special status.*

Once a color has been assigned a meaning, that meaning should remain consistent throughout all aspects of the interface. For example, if red is used to signal an alarm, all uses of red throughout the interface should denote the same meaning and user conclusions; whenever red appears, there's an alarm in the system. This creates consistency in the interface which helps users intuitively understand system interactions.

**[HFR-24]** **Implicit color associations of an accelerator complex should inform overall color usage in the digital control system interface. Secondary color associations that are common in everyday life and cultural applications should also be considered.**

*Details:* Implicit color associations of an accelerator complex should inform overall color usage throughout interface design. An example of this is when Fermilab accelerator structures are painted unique colors and those colors are replicated within the control system interfaces (e.g., the New Muon Lab building interior is painted mint green and the control system interfaces have the same color as a background). A better design for this example would be to include color associated headings (instead of background) to implicitly link accelerator infrastructure to relevant control system interfaces. *Technical Basis:* 0 (Section 6.3.1); 0 (Sections 7.2.5.1 – 7.2.5.10); 0 (Section 1.3.8)

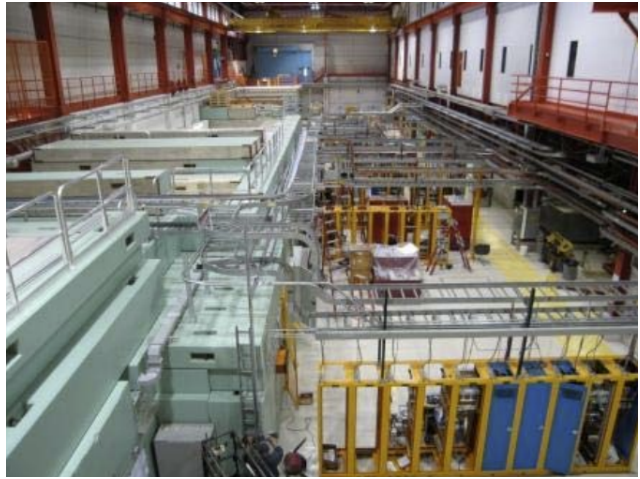

*Figure 10. New Muon Lab, interior paint color mint green.*

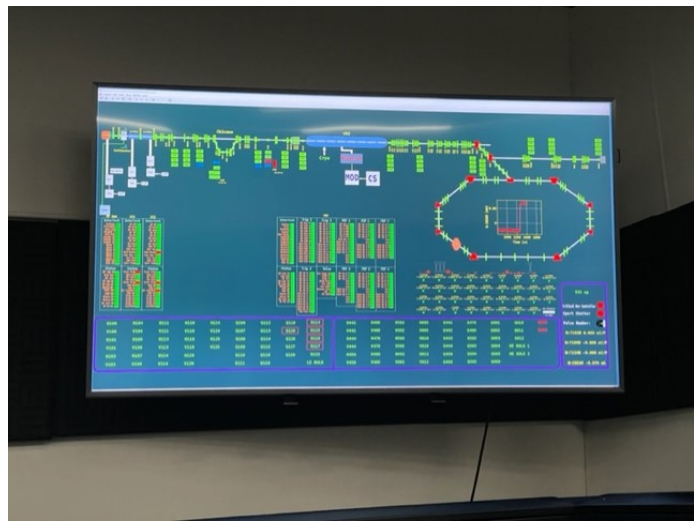

*Figure 11. New Muon Lab overview interface, background color mint green.*

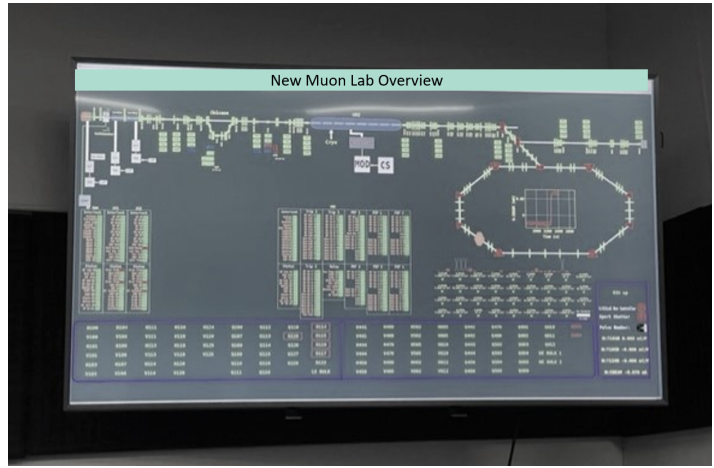

*Figure 12. Alternative design for implicit color association.*

**[HFR-25]** **A dull screen color scheme should be adopted to reduce display color saturation and saliency.**

*Details:* The "Dull Screen" approach is an interface design concept based on the theory that all normal behavior should appear "dull" so that abnormal behavior detected by the system can be highlighted or made salient through the use of color. 0. This strategy helps users rapidly detect events that require their detailed attention. This concept can also help reduce the amount of saturated colors included in an interface which improves a user's ability to differentiate between levels of information and focus on what is most important. *Technical Basis:* 0 (Section 6.3.1); 0 (Sections 7.2.5.1 – 7.2.5.10); 0 (Section 1.3.8)

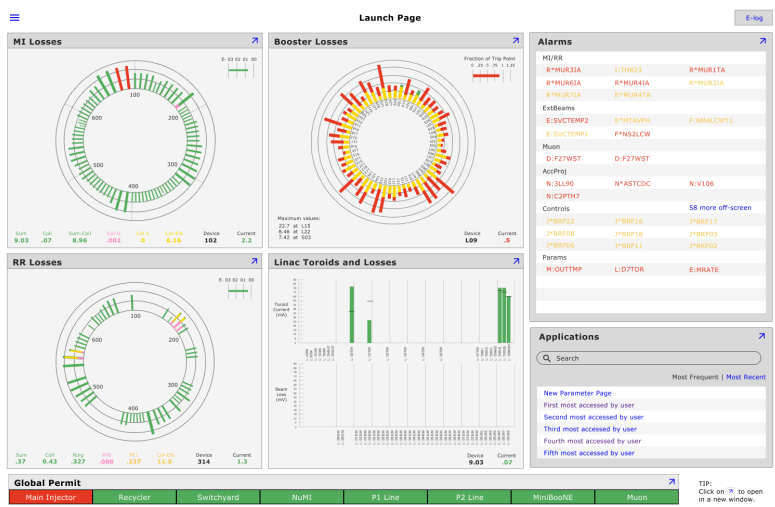

*Figure 13. Dull screen prototype (launch page).*

**[HFR-26]** **Saturated colors should be reserved to indicate special meaning.**

*Details:* Each color included in an interface competes with other colors and other display elements (e.g., text) for the user's attention. In line with the dull screen concept, highly saturated colors should be reserved for special or critical elements of a display. Special or critical elements of a display are those that must effectively draw a user's attention. In an accelerator digital control system, this can include multiple elements such as live data that must be continuously monitored or alarm events that require immediate attention. However, when too many elements on a display use color to convey special meaning, all information included in the display is disguised as equally important. In other words, if everything appears special, nothing appears special. Reserving saturated colors to indicate special meaning for certain display elements improves overall user performance and strengthens the comprehensibility of interface color-coding associations. *Technical Basis:* 0 (Section 6.3.1); 0 (Sections 7.2.5.1 – 7.2.5.10); 0 (Section 1.3.8)

**[HFR-27]** **Highest priority information (e.g., text or other display elements) must be tested for color blind safety.**

*Details:* Color blindness is the decreased ability to see color or differences in color within afflicted individuals. Color blindness affects about eight percent of males (approximately 10.5 million) and less than one percent of females. 0. There are two major types of color blindness: those who have difficulty between red and green, and those who have difficulty distinguishing between blue and yellow. A challenge in designing to accommodate color blindness is trying to accommodate the unknown. Not only do color-blindness types vary, but the level of color discernment ability varies between individuals as well. Although estimates of afflicted individuals are known, it is difficult to ascertain which color-blindness type potential users might have as well as what their level of color discernment is. Research models have been established to predict or calculate how colors are perceived by color-blind people, but they are not completely accurate. In other words, it isn't possible to predict future end user color blindness type and variability with 100% accuracy. Fortunately, there are some baseline recommendations that regardless of color blindness type and variability, will help accommodate color blindness.

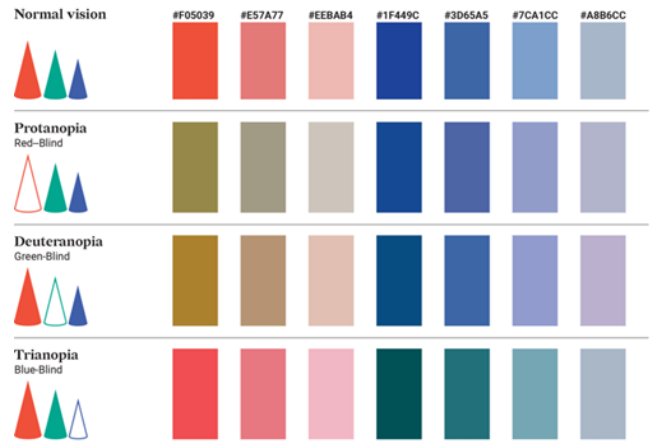

*Figure 14. Color blindness types and color combinations to avoid.*

The more important the interface content is, the more essential it is to make it color blind safe. Color blind safety is a concept that encourages certain types of color use to accommodate color blindness. The colors most detectable by anyone with color blindness are black and white (e.g., black text/elements on a white background and vice versa). This is because these colors have the highest contrast ratio compared to all other colors and are easily discernable from each other. If additional colors must be used, interface content areas should be monochromatic with the interface element color and background color at the opposite ends of the color saturation poles*.* Refer to the color palettes presented in Table 1 and Table 3. *Technical Basis:* 0 (Section 6.3.1); 0 (Sections 7.2.5.1 – 7.2.5.10); 0 (Section 1.3.8)

**[HFR-28]** **The system should apply the project defined color palette consistently across control system display pages.**
*Details:* All colors were selected from RColorBrewer® data visualization colorblind-friendly color palettes *(Appendix D, Figure 36).* A color-coding scheme reveals what category a display element fits in as each primary color conveys meaning to the end-user. A properly selected color-coding scheme enables quick identification of display information such as knowing a relevant category quickly without having to search to understand its contents first. *Technical Basis:* Refer to the color palettes presented in Table 1 and Table 3. *References:* 0 (Section 6.3.1); 0 (Sections 7.2.5.1 – 7.2.5.10); 0 (Section 1.3.8)

| Color | RBG Value & Hex Number | Primary Function | Description | Additional Information |
|---|---|---|---|---|
| Light Theme Colors | | | | |
| Red 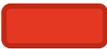 | R:229, G:56, B:35 Hex: #E53823 | Alarm, Danger Signal | Critical/High-Priority alarms | All types of faults that interfere with safety-critical operations |
| Orange1 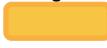 | R:248, G:196, B:67 Hex: #F8C443 | Alert, Caution Signal | Caution/Lower-Priority alerts | All types of faults that interfere with normal operations |
| Yellow 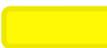 | R:255, G:249, B:5 Hex: #FFF905 | Alert, Acknowledge Signal | Acknowledge alerts | All types of alerts that operators must acknowledge |
| Green1 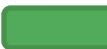 | R:82, G:171, B:90 Hex: #52AB5A | Live value output | Current live output of a numerical value | |
| Blue 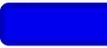 | R:0, G:0, B:238 Hex: #0000EE | Dynamic/Clickable element | Elements that are dynamic or clickable | |
| Purple 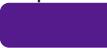 | R:85, G:26, B:139 Hex: #551A8B | Accessed dynamic element | Dynamic elements that have been previously clicked by user | Previously clicked elements should remain purple until user leaves page |
| Magenta 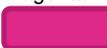 | R:222, G:40, B:138 Hex: # DE288A | Historical data | Elements that convey previous settings or historical data | |
| White 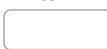 | R:255, G:255, B:255 Hex: #FFFFFF | Process lines Delineate objects | Provides perimeter around display elements White space | Secondary color for process lines and object delineation |
| Black 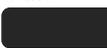 | R:38, G:38, B:38 Hex: #262626 | General text (static) Process lines Delineate objects | General text default Provides perimeter around display elements | |
| Light Gray 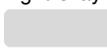 | R:217, G:217, B:217 Hex: #D9D9D9 | Background | Provides primary background color, light theme | |
| Gray 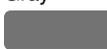 | R:150, G:150, B:150 Hex: #969696 | Static elements | Static display elements default | |

*Table 1. Primary Color-coding associations for accelerator control system (Light Theme).*

| Color | RBG Value & Hex Number | Primary Function | Description | Additional Information |
|---|---|---|---|---|
| Dark Theme Colors | | | | |
| Red 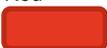 | R:229, G:56, B:35 Hex: #E53823 | Alarm, Danger Signal | Critical/High-Priority alarms | All types of faults that interfere with safety-critical operations |
| Orange1 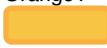 | R:248, G:196, B:67 Hex: #F8C443 | Alert, Caution Signal | Caution/Lower-Priority alerts | All types of faults that interfere with normal operations |
| Yellow 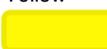 | R:255, G:249, B:5 Hex: #FFF905 | Alert, Acknowledge Signal | Acknowledge alerts | All types of alerts that operators must acknowledge |

| Green1 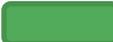 | R:82, G:171, B:90 Hex: #52AB5A | Live value output | Current live output of a numerical value | |
|---|---|---|---|---|
| Blue 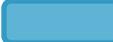 | R:95, G:180, B:213 Hex: #5FB4D5 | Dynamic/Clickable element | Elements that are dynamic or clickable | |
| Purple 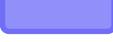 | R:145, G:143, B:249 Hex: #918FF9 | Accessed dynamic element | Dynamic elements that have been previously clicked by user | Previously clicked elements should remain purple until user leaves page |
| Magenta 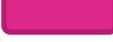 | R:222, G:40, B:138 Hex: # DE288A | Historical data | Elements that convey previous settings or historical data | |
| White 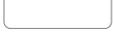 | R:255, G:255, B:255 Hex: #FFFFFF | Process lines Delineate objects | Provides perimeter around display elements White space | Secondary color for process lines and object delineation |
| Black 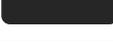 | R:38, G:38, B:38 Hex: #262626 | General text (static) Process lines Delineate objects | General text default Provides perimeter around display elements | |
| Gray 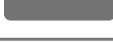 | R:150, G:150, B:150 Hex: #969696 | Static elements | Static display elements default | |
| Dark Gray 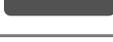 | R:82, G:82, B:82 Hex: #525252 | Background | Provides primary background color, dark theme | |

*Table 2. Primary Color-coding associations for accelerator control system (Dark Theme)*

| Color | RBG Value & Hex Number | Primary Function |
|---|---|---|
| Light Gray 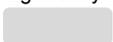 | R:217, G:217, B:217 Hex: #D9D9D9 | Primary background plot color, Light theme |
| Gray 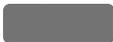 | R:150, G:150, B:150 Hex: #969696 | Secondary background plot color |
| Dark Gray 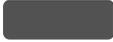 | R:82, G:82, B:82 Hex: #525252 | Primary background plot color, Dark theme |
| White 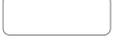 | R:255, G:255, B:255 Hex: #FFFFFF | Plot grid color |
| Green2 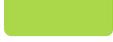 | R:170, G:216, B:74 Hex: #AAD802 | Plot color 1 |
| Gold 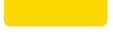 | R:251, G:216, B:0 Hex: #FBD800 | Plot color 2 |
| Orange2 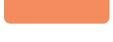 | R:244, G:140, B:93 Hex: #F48C5D | Plot color 3 |
| Pink 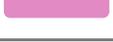 | R:225, G:138, B:196 Hex: #FF8AC4 | Plot color 4 |

*Table 3. Primary Color-coding associations for accelerator plots. Each color passes the color-blind test.*

Including color in interface design does not guarantee improved performance and user comprehension. Randomly adding or including colors that haven't been pre-approved and screened for accessibility (i.e., included in the palette) can inhibit performance and comprehension by causing visual clutter and confusion. Additionally, poor use of color across the entirety of an interface will diminish the effectiveness of colors conveying high priority information. This directly affects a user's ability to interpret the interface and respond to time sensitive information.

If developers suspect that existing colors aren't satisfactory and introducing a new color is justified, human factors/usability experts can review requests to introduce a new color into the existing palette. If the request is approved, the new color and descriptions will be incorporated into the style guide for future versions.

### 4.2.3  Text: Titles, Labels, Abbreviations, and Acronyms

**Font Size and Type**

**[HFR-29]**   **All alphanumeric text (static and dynamic) should be no less than 9-point font (or 16 minutes of arc) for adequate legibility.**

*Details:* A 9-point font yields approximately 16 minutes of arc when viewing at a seated workstation. Font sizes should be adjusted based on the intended viewing distance (Table 4); for example, text to be read at a distance of 10 feet (e.g., on an OVD) would require 39 point to achieve 16 minutes of arc on a 1920 x 1080 monitor. *Technical Basis:* 0 (Sections 7.2.6.1); 0 (Section 1.3.1-4)

| View Distance | Minutes of Arc | Font Height in Inches | Font Height in Pixels | Font Point Size |
|---|---|---|---|---|
| 24 inches | 16 | 0.11 | 11 | 9 |
| 48 inches | 16 | 0.22 | 21 | 16 |
| 62 inches | 16 | 0.29 | 27 | 21 |
| 84 inches | 16 | 0.39 | 37 | 28 |
| 120 inches | 16 | 0.56 | 52 | 39 |
| Based on 1920 x 1080 resolution. | | | | |

*Table 4. Font sizes by viewing distance.*

**[HFR-30]**   **All alphanumeric text should be Verdana.**

*Details:* Verdana is a sans-serif font that provides sufficient inter-character distinguishability (Figure 15). For instance, Verdana provides adequate distinguishability between characters for readability. *Technical Basis:* 0 (Section 1.3.1-2, 1.3.1-3), 0

Fermi National Accelerator Laboratory                                                                                                27

| Font Type | Uppercase I | Lowercase i | Uppercase L | Lowercase l | Number 1 |
|---|---|---|---|---|---|
| Verdana | I | i | L | l | 1 |
| Tahoma | I | i | L | l | 1 |
| Arial | I | i | L | l | 1 |
| Lucida Sans | I | i | L | l | 1 |
| Helvetica | I | i | L | l | 1 |
| Calibri | I | i | L | l | 1 |

*Figure 15. Readability comparison of common sans-serif fonts.*

An accelerator digital control system uses many letters and numbers in titles and labels. It is crucial for a user to be able to distinguish number "1" from letters "l" and "L." Verdana and Tahoma are both Sans-serif fonts where uppercase and lowercase "I," "L," and number "1" are easily distinguishable from one another.

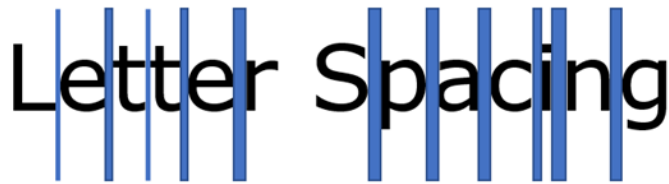

*Figure 16. Letter spacing for Verdana font type.*

Letter spacing (i.e., the spacing in between letters and numbers) should have enough room in between to distinguish letters without forfeiting the readability of the word. The letter space should always be smaller than the counterspace of the letter (i.e., the space of the letter itself).

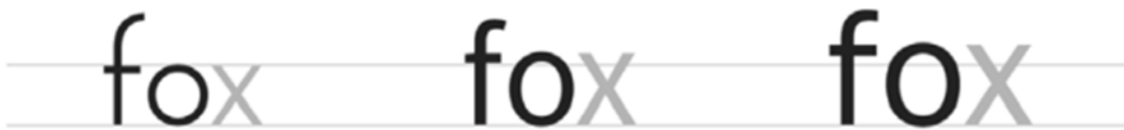

*Figure 17. Font x-height example 0.*

The x-height of a font type is an important feature because when it is accommodated properly, lowercase "x's" are easily distinguishable from upper case "X's." Confusing a lowercase "x" for an uppercase "X" can be confusing. Too large x-height causes difficulty in distinguishing additional letters such as *n* and *h (see Figure 17)*.

**[HFR-31]** **All alphanumeric text variations should be consistent throughout all interfaces.**

*Details:* Once a design convention is established, it should be consistently implemented throughout all interfaces and text formatting is no exception. Deviating from the standard text format convention can cause interface confusion and disrupt user workflow. For example, use commonly accepted formatting (e.g., capitalize first letter of titles) throughout all interfaces. *Technical Basis:* 0 (Section 1.3.1-2, 1.3.1-3), 0

**Labels**

**[HFR-32]** **The system should use a label convention that is intrinsically meaningful to the users.**

*Details:* Labels that are intrinsically meaningful to the user are an example of clear labels. For example, some of the main machines included in the accelerator complex are meaningfully named (e.g., the Booster machine "boosts" particles to increase their velocity). In addition to intrinsic meaning, clear labels also contain plain language that is concise yet descriptive (Figure 18). *Technical Basis:* 0 (Section 6.1.3); 0 (Sections 1.3.3-2, 1.3.3-5)

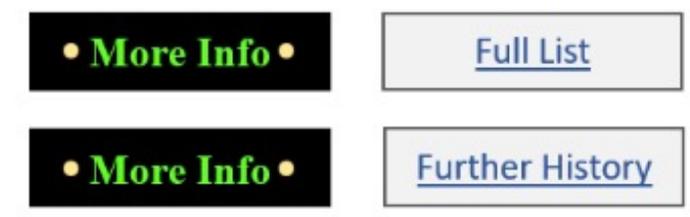

*Figure 18. Unclear label (left) and clear label (right; adapted from 0).*

Avoid labels and headings that are ambiguous and lengthy. An example of a clear and concise label is included below: The "more info" label is used to signify access to a complete list of digital statuses since not all statuses could fit on the display. It is also used to provide access to an expanded history of certain data parameters. "More info" is technically correct in both application; however, it isn't intrinsically obvious what "more info" means in either application. Alternative suggestions are represented on the right. These suggestions apply the clear label guidance of containing plain language that is concise yet descriptive, and it is immediately obvious what clicking on these labels will lead to.

**[HFR-33]** **The labeling convention should be applied consistently across the system.**

*Details:* Abbreviations, acronyms, mnemonics, and codes should be consistent within and between systems and applications that are anticipated to be used by the same user population. *Technical Basis:* 0 (Section 5.1.3.11.1.8.3)

**[HFR-34]** **All labels to be read should be oriented horizontally on display pages.**

*Details:* Labels and information (i.e., words and symbols) should be oriented so that alphanumeric characters are read horizontally from left to right. If labels are static in nature and secondary (e.g., a label for a y-axis plot that is static), then the label may be oriented vertically like in Figure 19. *Technical Basis:* 0 (Section 5.4.4.1)

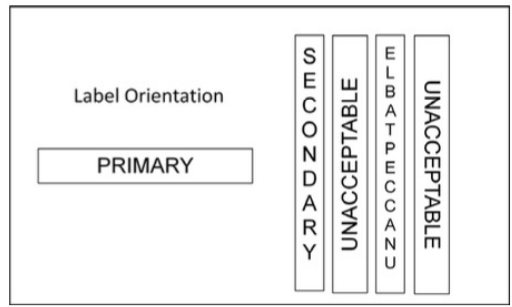

*Figure 19. Orientation of labels (adapted from 0)*

**[HFR-35]** **The system should visually differentiate between labels that are clickable and labels that are information only.**

*Details:* Visual element design can help provide at-a-glance context of information which yields more intuitive interpretation of content included in an interface. Multiple types of information are included in user interfaces including data that is interactive (i.e., links) and data that is information only. The purpose of visually differentiating between interactive data and information-only data is to provide at-a-glance context which information users can interact with. An example of how to differentiate between interactive and information only graphics is through visual elements that are intrinsically clickable. In the example included below, the text on the left is a clickable link and the text on the right is information only. The link text is not only a different color, but it is also underlined. Blue text that is underlined is a widespread design concept used for digital interfaces such as web pages. Therefore, humans have learned to associate blue, underlined text with links. *Technical Basis:* 0

*Figure 20. Link text versus information only text.*

*Figure 21. "Clicked" indication. The purple text on the right has been accessed recently.*

The purpose of visually displaying when linked information has been accessed recently is to inform a user that their control action has been received by the system. A control action can be something as small as accessing a link to something as large as shutting down the system. Regardless of the impact of the control action, users should be able to see at-a-glance that their action was received. This helps users efficiently interpret current information as well as previous information (i.e., recently accessed information). A simple but effective way to indicate that a control action has been initiated is to display a visible difference of the control/link before the action and after the action. This is a type of feedback that is very subtle and nondisruptive to normal operations but supports a user's ability to effectively operate a control system.

"Accessed recently" is a somewhat ambiguous phrase and is context dependent on the requirements of the control system. A good principle to apply if the requirements of the control system aren't known is to display the link as "accessed recently" (i.e., purple) until the action is completely initiated or until the user is no longer visiting that page. The primary purpose of distinguishing between regular links and links that have been recently accessed is to inform the user of the history of their own interaction with the system. At a minimum, the design should support the time it takes to convey that the user has recently accessed a link.

### 4.2.4 Dynamic Display Element Formatting

Graphics and images can be efficient to communicate great amounts of information with little to no text. Used properly they can help structure a page and act as an excellent resource for the user. Images in this context refer to pictorial representations of something that could include screenshots, drawings, and pictures of equipment, locations, or tools. Here, graphics refers to diagrams, mimics, simplified representations of equipment, or symbols. For the purposes of this document, the distinguishing characteristic between images and graphics is their complexity. Defining them as such helps to divide how they can function when designing an interface. See Figure 39 for an example of the difference in complexity and detail.

Plots are also discussed here and treated as their own category given the ubiquitous use of plots throughout accelerator control and monitoring but also the different types of plots used to assess the current system status of the accelerators.

**Trend Graph and Bar Chart Formatting**

**[HFR-36]   Clearly distinguish contextual information from live plot data.**
*Details:* Presenting operational context within plots (see figure 22) supports rapid recognition of system status. As adding operational context increases the information density of a plot, the design should make distinguishing the operational context from the live data easy to recognize. ACNET often displays new data over the old to help show change over time. However, there is little by way of design to distinguish the entry time of each new reading. Distinguishing the age of the reading in the plot may lead to some emergent features and understanding by users. *Technical Basis: 0*

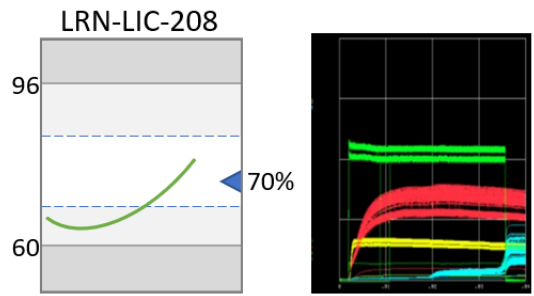

*Figure 22. The left figure shows live data (green, salient) against its operational context (light gray, neutral). The right figure shows data collected over many retrievals, but not indication of the timeline of retrieval.*

**[HFR-37]** **Present only necessary data on a plot to improve user time to complete task or understand system status.**

*Details:* Time to complete a task increases with task complexity. The more complex a task the more important it is to determine exactly what information is required. Tendencies to provide extra information can only increase the user need to process and interpret what is important and necessary from the extra information available. Developing a plot to fit a task will improve user's ability to perform the task. However, this pertains to novice users performing regular tasks. The capability to craft plots based on an emergent need should be available. *Technical Basis: 0*

**[HFR-38]** **Graphs and charts should include labels for its title, axes, parameters, and engineering units.**

*Technical Basis:* 0 (Section 5.17.21.1)

**[HFR-39]** **Graphs and charts should include a digital readout of the parameter(s) being represented when precise reading is required of the user.**

*Technical Basis:* 0 (Section 5.17.21.4)

**[HFR-40]** **Where multiple are presented on a single graph or chart, each parameter should be coded through the use of color or line type for differentiation.**

*Technical Basis:* 0 (Section 5.17.21.2)

**[HFR-41]** **Where multiple graphs and charts are compared on a single display page, their x-axes and y-axes should consist of the same scale.**

*Technical Basis:* 0 (Section 5.17.21.5)

**Mimic Display Formatting**

**[HFR-42]** **All components, line points, and termination points presented on a mimic display should be labeled.**
*Technical Basis:* 0 (Sections 1.2.8-2, 1.2.8-3, 1.2.8-4)

**[HFR-43]** **Flow direction should be coded with distinctive arrowheads.**
*Technical Basis:* 0 (Sections 1.2.8-5)

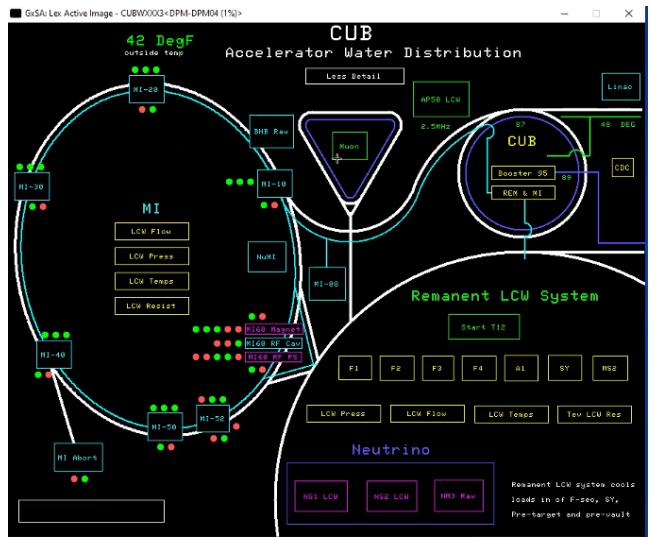

*Figure 23. Mimic diagram.*

## 4.3 Controls and Interaction

In the context of this style guide, a control is defined as an interaction that commands, directs, or regulates the behavior of physical devices on the system. The guidance in this section applies to aspects of the system where operator input directs action on the accelerator system.

Controls can be simple input fields where operators change values or set targets for the system to maintain (see Figure 24), or they can be software interfaces that mimic the look and feel of physical control devices (see Figure 25). The latter is often referred to as a soft control or a control faceplate. The guidance applies to both types. However, guidance in the visual representation section is more applicable to soft controls.

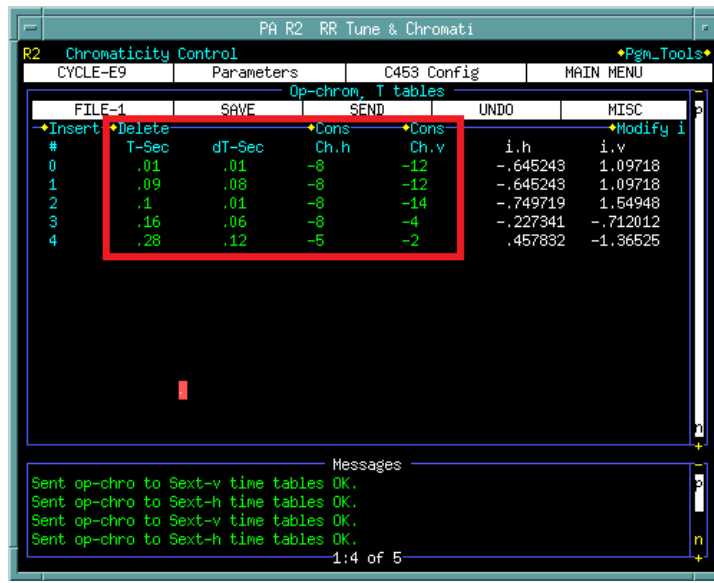

*Figure 24. Example of control input fields marked with the red box.*

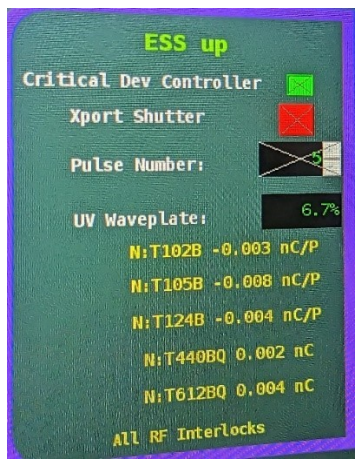

*Figure 25. Example of a control faceplate.*

### 4.3.1 User Input and Controls

Make it obvious where input or actions are possible and which elements of the interface are static or for informational purposes. For example, in Figure 24, there is not visual cue to indicate that the operator can edit the input fields. In Figure 25, the input fields for UV waveplate s denoted by a black box.  An alternative to the way Figure 24 is designed is shown in Figure 26, where all the input filed are denoted visually as input field and all the process data that is presented to monitor process feedback is presented as green text to denote that it is live data coming from the system.

|  | Page Title | | | | |
|---|---|---|---|---|---|
|  | Column Label | Column Label | Column Label | Column Label | Column Label |
| Row Label | ☐ | ☐ | ☐ | XXXXXXXX | XXXXXXXX |
| Row Label | ☐ | ☐ | ☐ | XXXXXXXX | XXXXXXXX |
| Row Label | ☐ | ☐ | ☐ | XXXXXXXX | XXXXXXXX |

*Figure 26. Example of design for a table with multiple input values.*

When controls have multiple options or modes (e.g., automatic, or manual) Make it clear what mode the controls are currently on and how that affects the functionality of the control (i.e., what inputs are available in a particular mode). For example, in the left side of Figure 27, Mode a is selected and there are only two control parameters that an operator can manipulate. The empty input field and the cursor indicate that no input has been selected. In the left side, mode B is selected, which has two input options. The second option is not shown when mode A is selected.

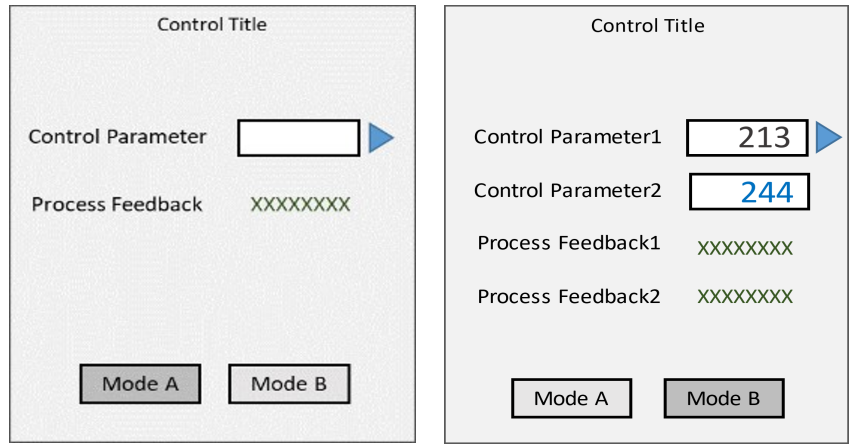

*Figure 27. Example of faceplate or soft control design.*

The visual design of controls should be consistent with the overall design. Especially with control faceplates it is tempting to mimic physical devices by adding 3D effects or texture to the design. This adds unnecessary visual clutter to the design and should be avoided.

**Input Devices**

**[HFR-44]** **The primary input device mode should be cursor-based (i.e., a mouse).**
*Details:* Navigation should be directed by mouse interaction (primary) and keystrokes (secondary). *Technical Basis:* 0 (Section 6.1.1)

**[HFR-45]** **The secondary input device mode should be keyboard-based (i.e., use of shortcuts for navigation).**
*Details:* The accelerator control system navigation should support flexibility for search-based navigation. *Technical Basis:* 0 (Section 6.1.1)

**Soft Controls**

**[HFR-46]** **The system should provide indication of all display elements that include control functionality.**
*Details:* There are current instances of "invisible" buttons on the screen that are used during operator tasks but provide zero indication of their presence or function. The new ACORN interface should clearly distinguish the function and presence of all control options available on screen. *Technical Basis:* 0 (Section 6.4)

**[HFR-47]** **All control options for a specific soft controller (i.e., faceplate) should be made accessible by a single click.**
*Details:* Control buttons (generally clicks) or inputs (generally alphanumeric entries) are those that operators use to adjust a component or system status. This includes but is not limited to adjusting setpoints, clearing alarms, navigating menus, and tuning machines. All control design should have the operational context for controlling, such as current setpoints, values, and predetermined thresholds, viewable while preparing to act. *Technical Basis:* 0 (Section 6.4.1.1); 0 (Section 7.2.1-1)

**[HFR-48]** **All frequency performed control actions should be accessible from a soft control faceplate without any additional administrative action.**
*Details:* All available or routinely used functions for control of equipment or components should be continuously visible or a maximum of one click away. *Technical Basis:* 0 (Section 6.4.1.5)

**[HFR-49]** **Soft control options should be suitable for characteristics of the task performed, by using Table 5.**
*Technical Basis:* 0 (Sections 7.2.4)

| Control Type | Example | Suitable Situations |
|---|---|---|
| Discrete Adjustment Controls | 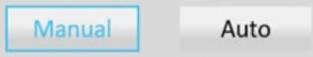 | • When control options are limited to a few. |
| Soft Sliders | 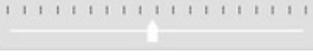 | • The range of possible values is medium-high.<br>• When changes are incremental and large (gross changes are needed). |
| Arrow Buttons | 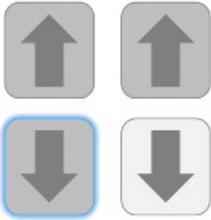 | • When precise value is needed.<br>• The range of possible values is medium-high.<br>• When changes are incremental and small. |
| Input Field | 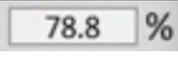 | • When precise value is needed.<br>• The range of possible values is high.<br>• When changes are variable (rather than incremental). |

*Table 5. Characteristics of different soft control types.*

**[HFR-50]** **Soft controls should be visually distinguishable from administrative buttons like navigation buttons.**
*Technical Basis:* 0 (Sections 7.3.1-5)

**Data Entry and Scripts**

**[HFR-51]** **Data entry should be accompanied by a verification step.**

*Details:* The system should begin processing a user's data entry input after pressing ENTER or selecting a confirmation button from the user interface. *Technical Basis:* 0 (Section 7.3.3-1)

**[HFR-52]** **The system should provide the capability of creating a control script for complex sequence of actions.**

*Technical Basis:* Key Insights (Insight 5A)

Error Prevention and Recovery

Provide clear explanation on why the input is invalid in alerts to the operator. If an operator were to enter a number that is too high, the alert would ideally state that the input is too high and state what the acceptable threshold is. In the event of the need for an operator to override a preset threshold, an alert should still appear in the same manner and include a confirmation that the operator would like to override.

**[HFR-53]** **The system should enable the user to correct for erroneous entries during data entry.**

*Details:* Where users are required to make entries into a system, an easy means shall be provided for correcting erroneous entries. *Technical Basis:* 0 (Section 5.1.2.1.7.1)

**[HFR-54]** **The system should provide confirmations for control actions that are safety important or have potential to disrupt normal operation.**

*Details:* Feedback is most crucial when a user is confirming actions that are safety related or for confirming actions that have the potential to disrupt normal operation (e.g., device shut down). *Technical Basis:* 0 (Section 6.4.2), 0 (Section 5.1.2.11.5.3)

**[HFR-55]** **The system should provide a means of restoring user defined settings in the event of a system failure.**

*Details:* Since beam requirements change so frequently due to experiments, operators rely on a system feature known as "restoring from a save" to avoid manually resetting beam requirements in the event of a system failure that results in erased data. The control system accomplishes this by capturing and storing reference values for accelerator devices only when an operator initiates the "save" functionality. This feature is often utilized when experimenters request a save after they've positioned all accelerator equipment the way they want it. However, operators shouldn't have to remember to save every time new beam requirements are revealed. *Technical Basis:* Key Insights (Insight 8A).

**[HFR-56]** **The system should prohibit multiple users from controlling the same equipment.**

*Details:* When asked which mistakes are most common, an operator stated that "two operators conducting the same task" (e.g., knobbing a machine) occurs frequently. Typically, it doesn't cause severe issues but sometimes does cause delays in operations. *Technical Basis:* Key Insights (Insight 7A)

### 4.3.2 System Interaction and Feedback

Provide feedback to identify:

Insights 1. What inputs are available, and what elements have been interacted with. Figure 27 illustrates how to denote what inputs are available by providing input filed for text (or numerical input) and buttons. Everything else is ither a title, a label or live process data. The feedback is provided by highlighting which mode is selected in dark grey. The top input filed has text that has been typed but not entered. And the bottom input filed shows text that has been input and entered.

Insights 2. Whether the system has carried out the requested action. Figure 27 shows that the system has accepted the input the operator provided in the second input field with blue text and by hiding the enter button until the operator enters new text into the field.

Insights 3. Where possible, what the system response is. Figure 27 shows how the process feedback can provide feedback on how the system response is for a related control parameter.

**System Response Time and Feedback**

**[HFR-57]** **Visual feedback should be provided across all user interactions with the system.**

*Details:* Every input by a user should produce a consistent, perceptible response output from the computer. *Technical Basis:* 0 (Section 5.1.2.1.6.3)

**[HFR-58]** **Visual feedback should be applied consistently across the control system.**

*Details:* The control system interface should support a variety of feedback interactions to inform users of autonomous system changes and user-initiated changes. Feedback messages should be consistent in style and format across the interface. *Technical Basis:* 0 (Section 6.4.2)

**[HFR-59]** **System latency should be 0.2 seconds or less for real-time responses.**

*Details:* Acceptable response times for key presses and error feedback is 0.2 seconds per Table 5 of 0. *Technical Basis:* 0 (Section 5.1.2.1.6.4)

**[HFR-60]** **The system should indicate that a user's input is processing for system response times greater than 1 second.**

*Details:* If computer response time will exceed 1 second, the user shall be given an indication (e.g., status bar or spinning wheel) or message indicating that the system is processing. *Technical Basis:* 0 (Section 5.1.2.1.6.4.3)

**Use of Blinking or Flashing**

**[HFR-61]** **Blinking/flashing should be used only for alerting the user to events that require immediate attention.**

*Details:* Flash coding should be employed to call the user's attention to mission-critical events only. *Technical Basis:* 0 (Section 5.17.27.1)

**[HFR-62]** **No more than two blink/flash rates should be used.**

*Details:* Flash rates should be limited to two types. They should be no more than 5 Hz and no less than 0.8 Hz; the difference should be greater than 2 Hz. The flash that is of higher importance should be of the greater frequency. *Technical Basis:* 0 (Section 5.17.27.3 & Sections 5.17.27.3.1 – 5.17.27.3.4)

**Alarms and Notifications**

**[HFR-63]** **Alarms should be used only for off-normal conditions that require timely action by the user.**

*Details:* The purpose of an alarm system is to direct the user's attention towards plant conditions requiring timely assessment or action. Each alarm should alert, inform, and guide the user. Every alarm presented to the user should be useful and relevant. Every alarm should have a defined response to which the user has adequate time to perform. *Technical Basis:* 0 (Section 1.3)

**[HFR-64]** **The system should provide a means to suppress alarms based on expected conditions from an experiment.**

*Details:* Different states in an experiment may influence alarms going off that are irrelevant to the experiment. The means to silence alarms that can be ignored at that time will allow the user to focus on the alarms that affect the experiment. *Technical Basis:* Key Insights (Insight 6A)

**[HFR-65]** **The system should provide the user with notifications of any conditions that may impact the accelerator's performance.**

*Details:* A somewhat surprising insight that was shared was how a variety of elements can influence accelerator devices. For example, the weather affects the way some magnets bend the beam. There isn't a direct measurement for this occurrence aside from posting the daily weather report in the control room which means operators must infer on their own why expected beam outputs are slightly off. *Technical Basis:* Key Insights (Insight 4A)

**[HFR-66]** **The system should provide an indication that the display is reading data from the control system (i.e., system heartbeat).**

*Details:* The user should know whether data is live or out-of-date. If data is not coming in from the system, this should be made clear to the user. *Technical Basis: 0 Insight 8*

# 5.0 HUMAN FACTORS FUNCTIONS-BASED STYLE GUIDE

## 5.1 Accelerator Displays

The organization and hierarchy of the control system is designed to support the tasks shown in Figure 28 (see section 7.0). The five tasks represent the workflow of main control room operators. Described within each task is the function performed for each. The design of the Human System Interaction is structured to facilitate the functions of each task. The accelerator display hierarchy is to be designed such that each display has the capability and functional support to aid an operator in performing their task comprehensively, quickly, and accurately. The following human factors requirements (HFR) call out what must be included in the accelerator displays to adequately support the functions within each of the five tasks.

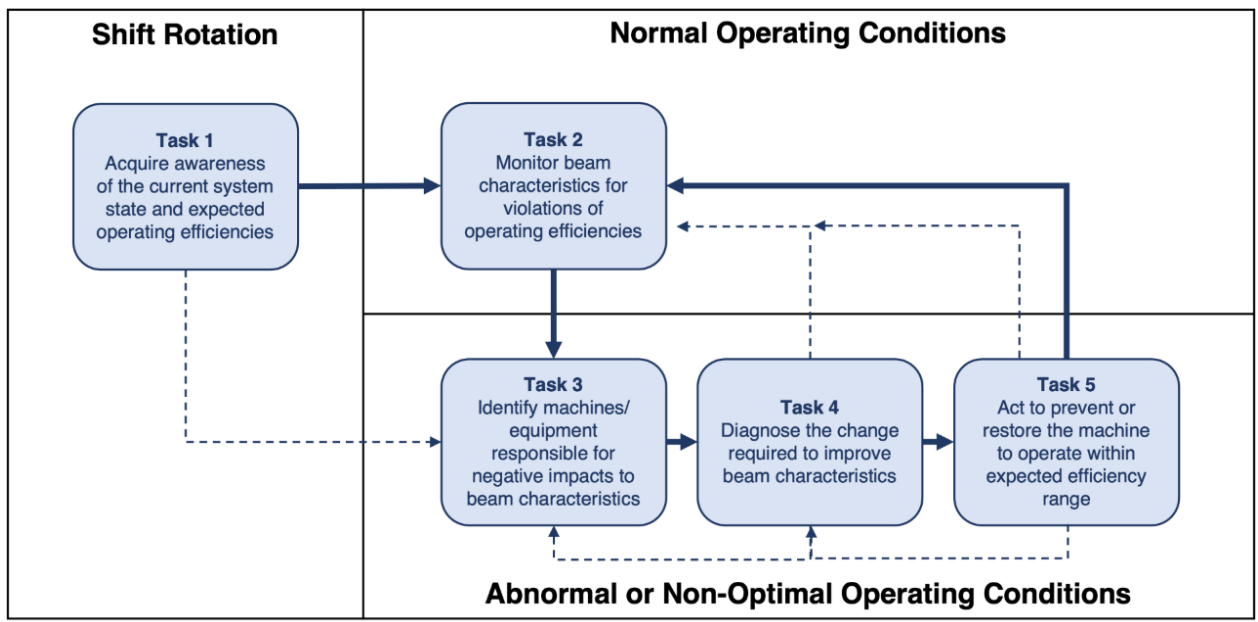

*Figure 28: A high-level task analysis depicting the general workflow of any operator maintaining beam quality of any nature.*

As stated, the human factors functional requirements are assigned to the different user interfaces that ensure that the human-system interaction support these functions safely, reliably, and efficiently within the user interface. While the interfaces will follow the task structure and support the associated functions, flexibility and maneuverability are key components of the overall display structure to enable ease-of-use and capability to all users that must interact with the accelerator control system.

The tasks and associated functions were derived from operator interviews, the results of which are synthesized in combination with the results from Appendix A (Section 6.0) and Appendix B (Section 7.0). While all interfaces will adhere to the global requirements of Section 4.0, this section will provide additional design requirement specifications (DRSs) that support HFRs derived from the task analysis. That is, the HFRs refer to 'what' should be accomplishable and the DRSs define 'how' it is accomplished through the display design.

### 5.1.1 Human Factors Functional Requirements for Monitoring, Detection, and Selection

The following HFRs support [HFR-4] specific to monitoring, detection and selection for accelerator displays (Tasks 1 (7.3.1) & 2 (7.3.2) Figure 28).

**Human Factors Functional Requirements to Support Awareness Acquisition and Monitoring**

**[HFR-67]** **The system should provide contextual status of 1) alarms 2) relative losses of critical machines 3) and access to an event summary report over the past shift to support shift briefings.**

*Details:* The typical shift rotation means an operator will spend about 16 hours away from operations before returning. Therefore task 1 is regaining awareness of what has happened in the past 16 hours and what is currently happening. This may include but is not limited to machines currently down for maintenance, problematic machines, status of alarm list, and a summary of what has occurred in the previous shift. *Technical Basis:* Task Analysis (Sections 7.3.1 & 7.3.2)

**[HFR-68]** **The system should provide status of available accelerator parameters deemed important for beam health monitoring across the accelerator complex.**

*Details:* Task 1 is for operators to acquire awareness of the current system state and expected operating efficiencies. Control room operators use a common set of plots to begin monitoring the beam health of different accelerator machines. The comfort displays that use the same plots emphasize the use of the common set of plots used to help operations maintain high-level awareness of accelerator system status. Typical indications include 1) main injector losses, 2) booster losses, 3) recycler losses, 4) beam budget liners, and 5) alarms. *Technical Basis:* Task Analysis (Section 7.3.1), 0 (Section 4.2)

**[HFR-69]** **The system should support the capability for a user to select available accelerator parameters from the control system that are deemed important for monitoring but not part of the standards MDS display loadout.**

*Details:* Currently, operators rely on 'comfort displays,' which are often redundant to the console windows but indicate what the crew chief has identified as important to monitor given the current accelerator statuses. These comfort displays are secondary sources in a nature and their effectiveness suffers due to the physical layout of the MCR that can restrict their visibility from some operator console locations. *Technical Basis:* Key Insights (Insight 4B), Task Analysis (Section 7.3.1)

**[HFR-70]** **The system should provide accessibility to the electronic logs (Elogs) for detailed descriptions of previous malfunctions, troubleshooting, maintenance, and issue resolutions.**

*Details:* Operators needing more detailed information about any part of the system consult the elog. Elog is a browser application and not constrained to the 5 window limit set by ACNET. Operators consult the elogs for detailed descriptions of previous malfunctions, troubleshooting, maintenance, and issue resolutions. The elogs can contain pictures, points-of-contact, and links to other helpful resources. *Technical Basis:* Key Insights Document (Section 6.3.1 3A), Task Analysis (Section 7.3.1)

**[HFR-71]** **The system should enable the user to monitor accelerator status and overall beam quality without any administrative task burden.**

*Technical Basis:* 0 (Section 1.1-14)

**Human Factors Functional Requirements to Support Monitoring, Detection, and Selection**

**[HFR-72]** **The system should notify users of machine/ equipment conditions that are indicative of negatively impacting beam characteristics requiring adjustment.**

*Technical Basis:* Operators are required to monitor emergent alarms and notifications. This is apparent in the "acquiring awareness of the current system state and expected operating efficiencies" section of the task analysis wherein typical windows used for everyday monitoring lists "alarm screen." *Reference:* [HFR-65]; Key Insights (Insight 4A)

**[HFR-73]** **The system should allow users to access detailed information of any abnormal machine/ equipment condition that requires adjustment.**

*Technical Basis:* Upon identifying an abnormal condition in the current system, the operator must close out windows that provide more holistic views of the system in favor of more detailed information regarding the inefficiency due to the five window ACNET limit. The operator also relies on knowledge gained through experience and training. The information provided by ACNET provides some historical data leading up to the inefficiency in question. It also provides at times provides the range of expected values for context. The operator is responsible for understanding what other factors may be impacting the inefficiency in question. *Reference:* Task Analysis (Section 7.3.3)

**[HFR-74]** **Displays should indicate when data presented is live or no longer up-to-date**

*Details:* Feedback on the use of a single MDS display to monitor overall system health and alarm statuses indicated that a display feature be added to show whether the on-screen data is live or stagnant. *Technical Basis:* Appendix A (Section 6.3.2)

**[HFR-75]** **The MDS page should support both proactive and reactive responses to accelerator statuses that may or currently require operator attention**

*Details:* Operators are responsible for monitoring for violations of expected beam quality. However, more skilled operators are capable of identifying "poor performers" and will improve the beam quality further. Therefore, operators should have information that allows for both proactive and reactive action to support both expectations for operators. *Technical Basis: Task Analysis, (Section 7.3.2).*

### 5.1.2 Design Requirement Specifications for Monitoring, Detection, and Selection

The following design requirement specifications (DRS) are the necessary elements of the interface that achieve the functional requirements of the monitoring, detection, and selection page (i.e., 'Launch Page'). These elements must also adhere to the global style guide requirements listed in Section 4.0.

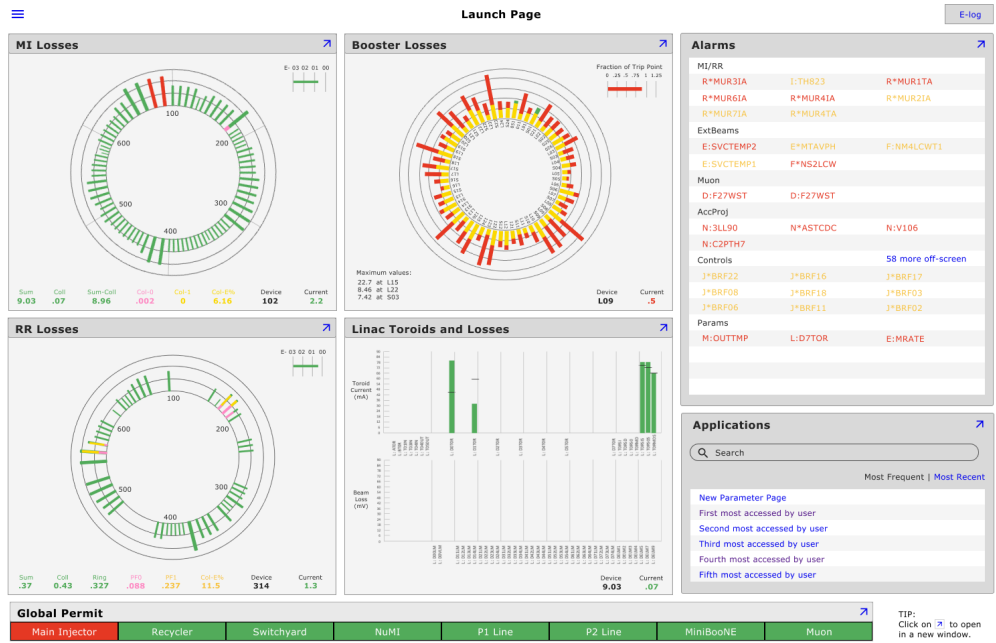

*Figure 29. Screenshot of the Launch Page (Light Theme).*

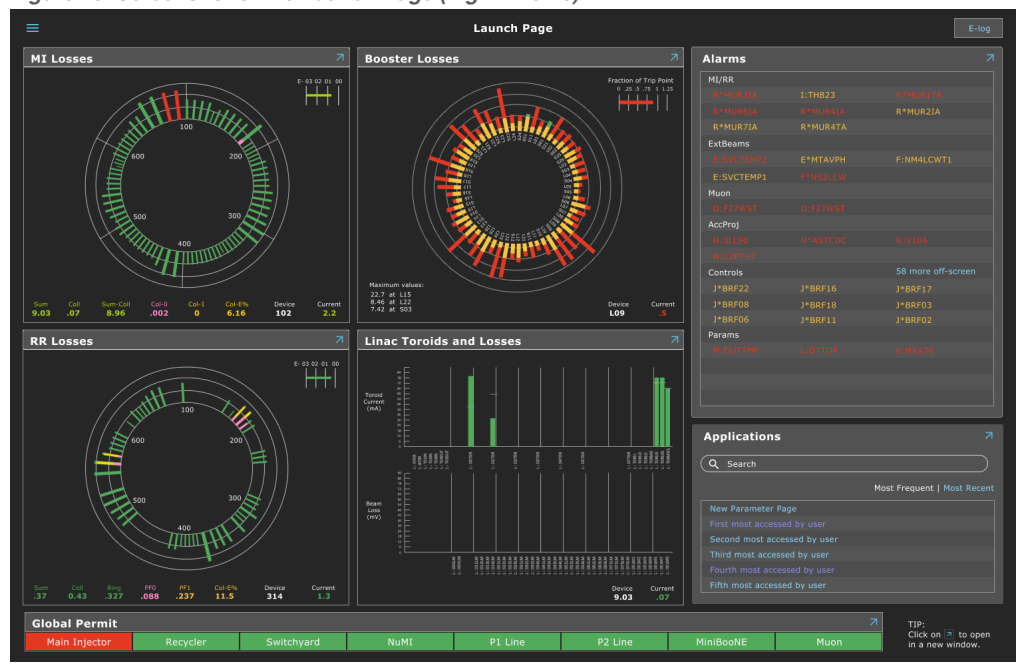

*Figure 30. Screenshot of the Launch Page (Dark Theme).*

HUMAN-SYSTEM INTERFACE STYLE GUIDE FOR ACORN DIGITAL CONTROL SYSTEMThe MDS page provides the following DRSs.

**[DRS-1]** **A single landing page should be designed for all users performing high-level monitoring of the accelerator complex.**
*Details:* The MDS landing page is designed to support monitoring, detection, and next-step selection of the accelerator complex and any indications that action must be taken (**Error! Reference source not found.**). Task 1 and Task 2 from Figure 28 should be accomplishable from the landing page with minimum interaction of interpretation. Therefore, the following DRSs must be met to ensure the landing page effectively supports the first two tasks in a user workflow. *Technical Basis:* [HFR-71]

**[DRS-2]** **Display a targeted set of dynamic, high-level information (such as plots with one or multiple parameters) for monitoring and detection purposes.**
*Details:* Key indications include but not limited to: main injector losses, booster losses, recycler losses, and beam budget monitor. *Technical Basis:* [HFR-67], [HFR-68], [HFR-69]

**[DRS-3]** **Display relative losses of key accelerator systems based on expected system operating parameters established at beginning of shift or after key activities.**
*Details:* The loss plots must indicate and distinguish between current losses acceptable for the current activities being performed, expected losses (e.g., collimators), decreased losses (I.e. improved efficiency) and key locations that impact the magnitude of losses or otherwise impact the users frame of reference for acceptable losses. *Technical Basis:*[HFR-55][HFR-56][HFR-59][HFR-63]

**[DRS-4]** **Display a system application list determined by user usage to efficiently navigate through the system for quick resolution and task completion.**
*Details:* User needs access to other apps and plots to accomplish tasks defined in task analysis. Quick actions can resolve beam events that require troubleshooting and/or adjustments. The fast response time minimizes the time a problem affects the beam. *Technical Basis:* [HFR-32][HFR-33]

**[DRS-5]** **Display priority alarms affecting beam to alert for operator response.**
*Details:* As stated in the "Key insights document" and the task analysis, operators are expected to respond (i.e., diagnose) to emergent alarms and notifications. This is apparent in the "monitor beam characteristics for violations of operating efficiencies" and "identify [machines/equipment] responsible for negative impacts to beam characteristics" sections of the task analysis wherein a primary source of operating efficiency violations are emergent alarms and responding to those alarms is the precursor to identifying the responsible machine/equipment. *Technical Basis:* [HFR-63][HFR-64][HFR-65][HFR-66]

**[DRS-6]** **Display search bar with function to locate any system application via name or identification code.**
*Details:* The monitoring and detection information on the landing page may indicate an abnormality that requires an application not on the users personalized application list. The search functionality allows a user to quickly locate the application that is needed for troubleshooting and solution. *Technical Basis:* [HFR-11]

### Human Factors Functional Requirements for Diagnosis and Acting

This section is out of scope for this revision of the style guide. Future revisions will include display specifications for these roles.

### 5.1.3 Design Requirement Specifications for Diagnosis and Acting

This section is out of scope for this revision of the style guide. Future revisions will include display specifications for these roles.

## 5.2 Analysis and Control Displays

This section is out of scope for this revision of the style guide. Future revisions will include display specifications for these roles.

## 5.3 Custom Parameter Page Displays

This section is out of scope for this revision of the style guide. Future revisions will include display specifications for these roles.

## 6.0   APPENDIX A – KEY INSIGHTS

## 6.1  Scope

The key insights described here serve as an input to the technical bases for the requirements described in this document, as part of Revision 1.

## 6.2  Introduction

This document details findings from an initial round of human factors interviews with accelerator operators at Fermilab within a consolidated file instead of raw data files (e.g., individual interview notes). The main findings are organized as key insights and a summary of associated design recommendations. Additionally, this document includes appendix references such as a glossary of accelerator terminology and a task analysis of operations at Fermilab.

### 6.2.1  Interview Protocol

Over the course of multiple months, the INL team met with 15 Fermilab main control room (MCR) operators to complete a semi-structured interview. The interviews were conducted to accomplish two goals: (1) to acquire operator specific knowledge about the accelerator control system and more importantly (2) to elicit operator feedback regarding their overall workflow including pain points.

The interviews took place over a virtual meeting platform. Most operators had the capability to share their screen to demonstrate their responses using ACNET, the primary control system at Fermi, as well as other tools typically used by operators. Each operator was asked to provide some demographic information about their educational background and operations experience at Fermilab.

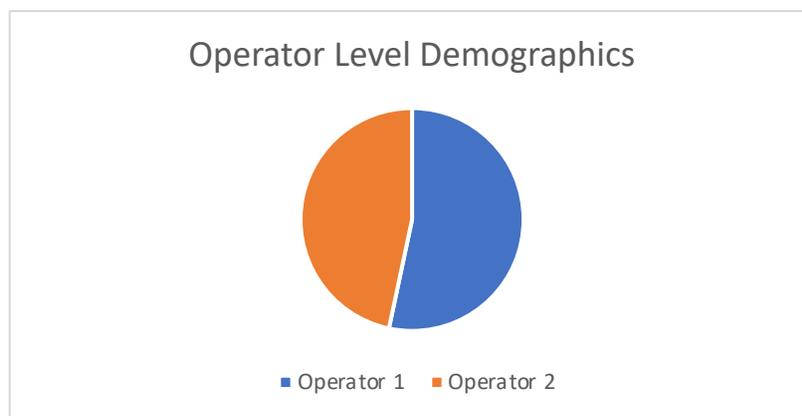

*Figure 31. Operator level demographics for key insights.*

There are two classifications of main control room operators: "operator 1" and "operator 2." The main difference between these classifications is operator 1s are considered new and require supervision to conduct their work whereas operator 2s are more experienced and can conduct work without supervision. Operator 2s are former operator 1s that have completed the operator training and passed the operator test which is required to advance. Of the 15 operator participants, eight were classified as operator 1s and seven were classified as operator 2s (see figure 1). The average years of operations experience between the participants was 2.32 years. Additionally, each operator had some sort of physics bachelor's degree, (i.e., engineering physics, nuclear physics, or experimental physics).
Interview questions focused on how operators perform their day-to-day duties and were structured to include four main categories:

- Data and information gathering
- Controls and adjustment making
- Crew dynamics and teamwork
- Available/needed support for operators

The semi-structured nature of the interviews afforded the flexibility for researchers to delve further into specific topics to clarify assumptions about certain operating tasks, gain new perspective on a typical daily task, and further explore the methods employed by any one operator. The flexibility became useful as some topics became stale, hearing the repeated explanations on a topic, and researchers could amend the list of questions to continue pulling new or further enriching information from operators as researchers gained familiarity with the ACNET system.

**Low Hanging Fruit.** Throughout the interviews, operator insights were captured and tallied (see Main Insights section). Some of these insights resulted in identification of design recommendations that would have an immediate and significant impact on control system improvements. These design recommendations are referred to as "low hanging fruit" and are captured in general guidance in [1] and more specific guidance in [2]. However, it should be noted that the purpose of these documents (ref 1 & 2) is to provide comprehensive design guidance for the development of accelerator interfaces. In other words, a succinct summary of "low hanging fruit" design recommendations regarding the first round of operator interviews is not included in either reference. Therefore, an additional purpose of this document is to not only summarize key insights, but also to provide design recommendations based on those insights to address the "low hanging fruit."

### 6.2.2 Accelerator Control System
Since one of the primary purposes of these interviews was to provide researchers an opportunity to acquire knowledge of the accelerator control system, this section includes brief descriptions of what the accelerator control system entails, how it is operated, and additional tools that support it.

The high-level purpose of the accelerator control system is somewhat self-explanatory; to control the accelerator. However, a more human/operations centric definition would be to grant front-end access of accelerator equipment (i.e., machines and hardware devices) to accelerator personnel to manipulate said equipment according to stated criteria. The accelerator control system includes thousands of devices (e.g., magnets, sensors, etc.) all of which contain some degree of control functionality. The primary tool utilized by accelerator personnel to control is the Accelerator Control Network (ACNET) which is a connectionless, peer to peer networking protocol used for front ends [3, controls rookie book]. Other tools integrate with ACNET, such as web-based applications, to compile the accelerator control system and provide accelerator personnel the ability to monitor, manipulate, and experiment with accelerator data.

### 6.2.3 User Roles

There are multiple types of roles that interact with the accelerator control system including operators, machine experts, and physicists/engineers. Each of these roles has somewhat unique goals and responsibilities concerning the accelerator control system.

- Operators serve as the first line of defense for detecting and responding to beam line abnormalities. Their primary purpose is to monitor and maintain the beam output as optimally as possible.
- Machine experts serve as equipment specialists for a certain device or a slew of devices. When operators are not able to properly diagnose an event, they defer to machine experts to troubleshoot and restore the beam.
- Physicists/engineers perform experiments and monitor experimental data. They make requests or initiate changes to better meet their experiment requirements.

Each of these roles interact with the accelerator control system for specific purposes, and although each role is important, the initial round of interviews focused solely on main control room operators. This means the documented insights included in this report only reflect operator feedback and not machine experts or physicists/engineers. However, it is the intent of the research team to conduct additional interviews to capture insights that represent all roles as a future effort.

## 6.3 Key Insights

These sections include a list of key insights gathered throughout the operator interviews. The purpose of these insights is to demonstrate knowledge elicitation of operator interaction and workflow of the accelerator control system as well as linking these insights to "low hanging fruit" design recommendations.

### 6.3.1 Insights without recommendations

| Insight 1. | General Operations |
|---|---|

**1A:** Many operators described their primary role in the control room as being responsible for the following: maintaining beam quality (i.e., keep things running well), troubleshooting when necessary, and identifying/notifying the proper person (i.e., machine expert) to contact when operator diagnosis is unsuccessful. When asked to quantify the frequency of operator diagnosis success rate, one operator estimated that 75-80% of the time, operators can diagnose and solve an issue almost immediately. Sometimes they require more time to diagnose but are still able to handle it themselves, but occasionally they must contact machine experts and hand off the issue for them to solve. A follow up question revealed that there are typically 4-8 experts available at any time to help operators diagnose and solve incidents.

**1B:** Additional insights regarding general operations include how beam requirements are defined. For example, some experiments change beam requirements as often as weekly. The typical protocol for how changes in beam line requirements are handled are as follows: beam line physicists position the beam line where they need it to be according to the requirements of their experiments and then operations is responsible for maintaining the newly defined beam line.

| Insight 2. | Tuning |
|---|---|

**2A:** Tuning is a frequent control behavior exhibited by operators. The purpose of tuning is to manipulate specific accelerator equipment to optimize beam quality (beam quality is sometimes subjective and varies by experiment). When asked about typical tuning behaviors, many operators stated that most expected tuning behaviors occur when an experimenter requests a tune or if an alarm goes off. However, if neither of those things happen, it's up to the discretion of the operator to determine how fine-tuned instruments should be. Another operator stated that each operator makes a lot of judgement calls when it comes to tuning, and they're pickier with their tuning when things are slow (i.e., no events to respond to) during shifts.

| Insight 3. | E-log |
|---|---|

**3A:** E-log is a repository of operator shift records. Multiple operators mentioned how it is common for them to access and scan the e-log for a variety of tasks spanning from tuning to troubleshooting. E-log provides recent event logs as well as descriptions of unique and historical event diagnosis. E-log is an isolated control application (i.e., separate from ACNET) that operators rely heavily on to provide relevant context to their everyday operations.

### 6.3.2 Insights with recommendations

| Insight 4. | General Operations |
|---|---|

**4A:** A somewhat surprising insight that was shared was how a variety of elements can influence accelerator devices. For example, the weather affects the way some magnets bend the beam. There isn't a direct measurement for this occurrence aside from posting the daily weather report in the control room which means operators must infer on their own why expected beam outputs are slightly off.

**Recommendation:** provide a notification to operators when outside temperatures are approaching parameters that are known to affect beam quality instead of relying on an operator's ability to automatically assume the weather is the source of beam deviations.

**4B:** An additional stated insight was console windows are limited to displaying a maximum of five plots. Each operator that mentioned this were asked a follow up question of "why?" and to their knowledge, there was no known reason but suspected it might be a hardware constraint. This constraint has been known to disrupt their workflow in some monitoring tasks, but mostly in diagnosis tasks.

**Recommendation:** if possible, remove the five-plot maximum window console constraint to allow operators more freedom in monitoring and manipulating accelerator data.

| Insight 5. | Pain Points |
|---|---|

**5A:** When asked about pain points of the accelerator control system, multiple operators mentioned how the index pages in ACNET are cumbersome to navigate and cluttered. For example, the programs of index pages are organized alphabetically instead of being prioritized in place of function or frequency of use. Additionally, some programs are obsolete but are still included which causes a lot of unnecessary visual clutter.

**Recommendation:** incorporate a way to prioritize index page programs according to place of function or frequency of use. Additionally, automate a way to eliminate obsolete programs or alternatively, conduct an "obsolete" sweep annually to manually delete obsolete programs.

**5B:** Another consistent pain point stated by operators is the concept of "blowing away plots." Blowing away plots happens because anytime an operator wants to refresh a plot (e.g., view most recent data), the previous data disappears and is replaced by the most recent data. One operator described it as follows, "there are secret buttons, like secret blocks in Mario, that allow [an operator] to refresh the plot but keep the existing data, but if [an operator] doesn't know where that secret button is, they lose that data when they refresh the plot. First, there is no reason to hide necessary control functions, it only increases the overall cumbersomeness of a system. Second, since operators have stated a need to view historical data and most recent data in conjunction, there should be functionality to support that.

**Recommendation:** make all invisible functionality visible. Provide an ability to automatically view historical data and most recent data simultaneously without erasing one or the other.

| Insight 6. | Alarms |
|---|---|

**6A:** Alarms play a unique role in accelerator operations. Current alarm functionality includes providing operators a check list of corrective actions as well as maintaining special operating conditions (i.e., bypassing alarms). However, multiple operators stated that the responsibility of remembering which alarms should be in bypass position and which should be active falls on operators. Additionally, there's not a proper resource to validate which alarms should be in which position because that can change often depending on current experiments. If an operator must bypass many alarms to meet experiment requirements, they will typically create a personal alarm list to help them remember and as a source of validation when needed.

A specific example shared was when one operator bypassed an alarm and forgot. Only when they noticed a system irregularity did they remember to switch alarm position. The consequence of this error can vary and result in minor delays or in more extreme situations, can cause beam failure.

**Recommendation:** replace the alarm bypass system with customizable alarms instead. This way, the operator can outsource maintaining special alarm conditions to the system.

| Insight 7. | Common Errors |
|---|---|

**7A:** Operators are highly trained and extremely skilled at what they do. However, when asked which mistakes are most common, an operator stated that "two operators conducting the same task" (e.g., knobbing a machine) occurs frequently. Typically, it doesn't cause severe issues but sometimes does cause delays in operations.

**Recommendation:** integrate permissions functionality to prohibit multiple users from controlling the same equipment at the same time.

| Insight 8. | Restoring from a Save |
|---|---|

**8A:** Since beam requirements change so frequently due to experiments, operators rely on a system feature known as "restoring from a save" to avoid manually resetting beam requirements in the event of a system failure that results in erased data. The control system accomplishes this by capturing and storing reference values for accelerator devices only when an operator initiates the "save" functionality. This feature is often utilized when experimenters request a save after they've positioned all accelerator equipment the way they want it. However, operators shouldn't have to remember to save every time new beam requirements are revealed.

**Recommendation:** Auto-saves should happen automatically each time new beam requirements are set unless an override is manually executed by an operator. Previous save records should be available when desired in case experiments are repeated or previous reference values are needed.

| Insight 9. | Flexible Operations |
|---|---|

**9A:** The troubleshooting and problem solving that occurs within accelerator operations is highly variable. One operator stated that there is a strong need for custom pages (i.e., a flexible program) because diagnosing and solving a unique problem is accomplished differently almost every time. However, an abundance of custom displays leads to visual clutter (see insight 5A) and therefore the need to be disciplined in display quality and review is stated.

**Recommendation:** Flexible operations (i.e., custom pages) should be supported with the caveat that quality control and review is mandatory.

| Insight 10. | Requested Functionality |
|---|---|

**10A:** One operator stated that the scattered nature of the accelerator control system (i.e., isolated applications) prolonged their ability to understand and become proficient in operating. An example of this was mentioned that when operators are tasked with monitoring main injector, the "main injector kickers" are located in a website URL that is completely separate from ACNET. The separation of related information was stated as difficult to overcome and a preference for consolidated information was detailed.

**Recommendation:** evaluate ways to centralize all necessary resources into one location within the control system.

### 6.3.3 Frequency of Insights

Since the interview format was semi-structured, the insights collected across interviews varied. However, a key insight was mentioned in all interviews.

Every single operator described their on-the-job experience as their primary resource of interpreting control system information. In other words, an abundance of tribal knowledge is required to effectively operate the accelerator control system. This causes operators to devote a large portion of their mental workload to learning and remembering unique intricacies of the system whereas they could be devoting that energy to more appropriate tasks such as alarm response. Addressing some of the "low hanging fruit" design recommendations will help reduce operator cognitive burden by eliminating cumbersome functionality, visual clutter, and error prone design.

## 6.4 Conclusion and Disclaimer

First and foremost, this document is intended to act as a resource of consolidated findings from the initial round of human factors interviews with accelerator operators at Fermilab. Additionally, this document details a summary of design recommendations to address the current state of the accelerator control system (more specifically ACNET) at the time when interviews were conducted. This document also includes additional resources such as a glossary and an operator task analysis. Although this document is not a comprehensive design guidance, and should not be referenced as one, it is a resource of consolidated findings and documentation of knowledge elicitation.

Incorporating insights from the first round of interviews is crucial to overall project success. However, not only do additional interviews for other accelerator personnel roles need to be conducted, but additional operator interviews need to occur as well. Therefore, a proposed follow-on activity is to develop a gap analysis of operator topics and questions to address with another round of interviews.

# 7.0 APPENDIX B – TASK ANALYSIS FOR SHIFT OPERATORS

## 7.1 Scope

Another outcome of the interviews was a high-level operator task decomposition. The task decomposition was developed through inferences taken from each interview and two interviews to validate the accuracy of those inferences. Main control room operators have the high-level goal to maintain beam quality within health, safety, and experimental parameters. The task decomposition targets the relative task sequence operators move through to carry out this objective. There is nuance to the sequence of tasks depending on the shift and accelerator state. However, for the bulk of operation the five listed tasks are sequentially executed.

## 7.2 Summary of Performed Tasks

Main control room operators at Fermi National Laboratory are charged with maintaining the quality and efficiency at which the various machines are providing and directing 'beam'. MCR operator's foremost goal is maintaining beam within the constraints of safety codes, to protect the welfare of those working at the lab, and equipment specifications to protect the machines providing and directing the beam. Within those constraints are the various experiment parameters that inform what the beam composition and quality must be. To operate within these parameters MCR operators carry out the following five tasks to perform their duties.

1. Acquire awareness of the current system state and expected operating efficiencies.
2. Monitor the various machines within the system for violations of operating efficiencies.
3. Identify machines at risk of or currently violating expected operating efficiencies.
4. Diagnose what is causing the potential or current violation of expected operating efficiency
5. Act to prevent or restore the machine to operate within expected efficiency range.

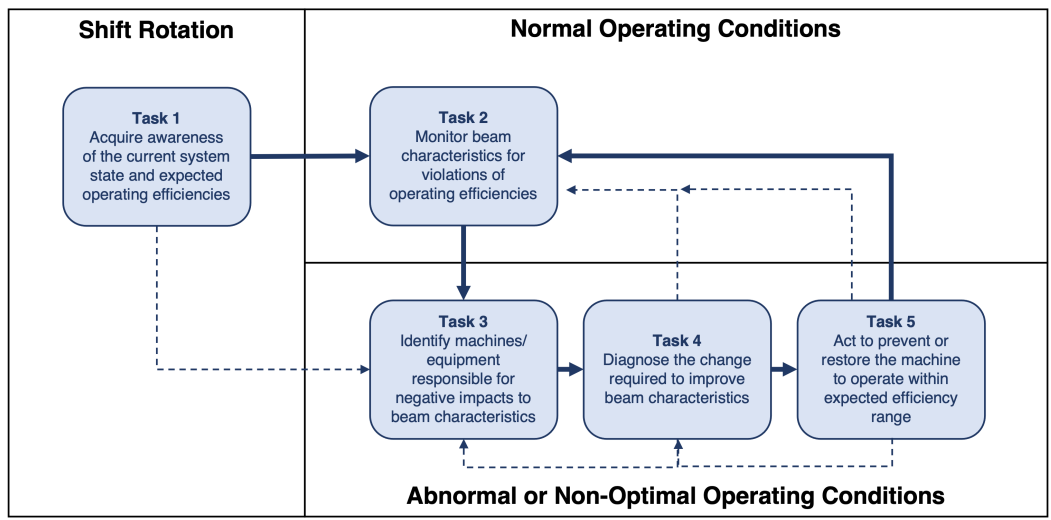

*Figure 32. High-level tasks performed by shift operators.*

The high-level tasks are cyclical in that once task five is accomplished the operator returns to the second task and continues in that order until a shift is over. At which point the exiting operators support incoming operators with the first task. The tasks at this level are successive in nature in that latter goals are built on the information gathered and actions taken in previous goals. System upgrades that support actions allowing operators to efficiently move from one task to the next will support the upgrade objectives.

During a shift, operators interact with ACNET, crew members, and browser applications to help them complete each task. Therefore, each high-level task can be broken down by the necessary actions or information needed for the operator to move from one high-level task to the next. The information requirements, coordination with crew members or other outside organizations, and specific task sequences operators typically adhere to are described in this task analysis.

## 7.3 Detailed Task Analysis

### 7.3.1 Task 1. Acquire Awareness of the Current System State and Expected Operating Efficiencies

The typical shift rotation means an operator will spend about 16 hours away from operations before returning. Therefore task 1 is regaining awareness of what has happened in the past 16 hours and what is currently happening. This may include but is not limited to machines currently down for maintenance, problematic machines, status of alarm list, and a summary of what has occurred in the previous shift. This orientation process helps operators know where to identify tasking during their shift. It also serves as a filter for operators to perhaps disregard some information in the control room that is otherwise used to identify tasking. One example is disregarding some known alarm statuses that are already being dealt with outside the control room. Operators without this information may attempt to act on a machine that is purposefully offline without knowing.

Two formal tasks carried out to inform incoming operators of the system state are 1) the briefing from their control chief occurring outside the control room and 2) a debriefing from operators of the previous shift. The crew chief briefing is a summary report of how the accelerators have been operating and any pertinent or safety-related information operators must know prior to beginning their shift. Then the previous shift hands off the controls to the new shift by presenting a very detailed accounting of what has happened and needs attending. Depending on the state of the system, this could mean directing new operators straight to task 4 or 5 in fixing a machine that is out of alignment.

Once the shift has begun operators use three different resources to acquire and maintain system awareness. First, is a set of windows that provide system wide performance information.  Operators typically set up the same five ACNET windows to orient themselves with the system they are operating.

Those windows typically include:
- Main injector losses plot
- Booster losses plot
- Recycler losses plot
- Beam budget liners plot
- Alarm screen

Second are the "comfort displays" that act as large overview displays. The comfort displays are often redundant to the console windows but indicate what the crew chief has identified as important to monitor given the current accelerator statuses. These comfort displays are secondary sources in a nature and their effectiveness suffers due to the physical layout of the MCR that can restrict their visibility from some operator console locations.

Third, is the highly detailed information that accounts for almost every action ever taken in the control room, the electronic logs, or elogs. The elogs are where all operators report and track their actions if a machine has received any kind of attention. Operators needing more detailed information about any part of the system consult the elog. Elog is a browser application and not constrained to the 5 window limit set by ACNET. Operators consult the elogs for detailed descriptions of previous malfunctions, troubleshooting, maintenance, and issue resolutions. The elogs can contain pictures, points-of-contact, and links to other helpful resources. After the briefings and console organization the operator typically has the information required to begin the second high-level task, monitoring beam characteristics for violations of operating efficiencies.

### 7.3.2  Task 2. Monitor Beam Characteristics for Violations of Operating Efficiencies

The second high-level goal is monitoring for operating inefficiencies. An inefficiency could be a violation of some beam characteristic that has moved outside the provided experiment, equipment or safety parameters. However, more experienced operators report they also seek out "poor performers" that have not yet violated a parameter threshold but could be made more efficient. Monitoring for inefficiencies requires filtering known statuses and expectations of the system with currently reported system status as presented in ACNET. It is possible some usual indications of violations can be ignored due to bypassed alarms, equipment that is moved offline or other reasons. Hence, acquiring system awareness is important before beginning to monitor beam characteristics. Monitoring tasks occur in no specific order. They include:
- Monitoring personally curated console windows
- Monitor beam profiles
- Monitor power supplies
- Monitor large view displays referred to as "comfort screens"
- Receive assignments to watch machines currently acting less predictably

- Receive calls from Experimenters to tune
- Compare current plot values with beam specifications from experiments

These tasks combine to create the high-level task of monitoring beam characteristics for inefficiencies. Performing these tasks create the opportunities necessary to move to the third high-level task; Identify machines responsible for negative impacts to beam characteristics. Causes for inefficiencies can vary. Often, the tasks in this section are enough to even identify the machines that need tuning; however, some situations require further diagnosis. Those steps are described in the following high-level task.

### 7.3.3 Task 3. Identify Machines and Equipment Responsible for Negative Impacts to Beam Characteristics

The third high-level goal is activated by two situations. 1) The operator may get an indication that a violation has occurred or 2) the operator recognizes a beam characteristic that could be improved. The operator then identifies first the machine where the inefficiency exists. Then identifies the equipment within that machine that is impacting the beam quality.

The information from the second high-level task is the initial input to determining an adjustment is needed. Once a potential adjustment is identified the operator essentially answers the following questions:

- Is the machine linear or cyclical?
- Is there equipment in this machine that has been reported as acting unusual?
- Is someone working on this piece of equipment already causing fluctuations in its efficiency?
- Could the beam quality here be affected by factors upstream of the identified inefficiency?

The first three questions can be answered by accessing information in ACNET such as plots, parameter pages, and downtime logs or browser applications such as the elog, schematics, and training tools. However, at this stage the operator must close out windows that provide more holistic views of the system in favor of more detailed information regarding the inefficiency due to the five window ACNET limit. The fourth question is answered using operator knowledge gained through experience and training. The information provided by ACNET provides some historical data leading up to the inefficiency in question. It also provides at times provides the range of expected values for context. The operator is responsible for understanding what other factors may be impacting the inefficiency in question.

The loss charts operators monitor indicate the machine in which an efficiency exists. Once identified, some diagnostic work is required to identify the right fix for the inefficiency. In general, linear machines are simpler to diagnose because equipment adjustments have only downstream effects. Therefore, beginning at the piece of equipment where the inefficiency is indicated is considered good practice. Cyclical machines can be more challenging as changes to one piece of equipment can affect the beam on either side of the equipment being tuned.

### 7.3.4 Task 4. Diagnose the Change Required to Improve Beam Characteristics

Once the machine is identified the operator completes the fourth high-level task by comparing the values on the parameter page related to the plot window associated with the equipment responsible for lack in beam quality. Often, the operator will make small adjustments to equipment settings and evaluate the result. If the plot values are moving in a favorable direction, the operator continues to the fifth high-level goal continuing to adjust until the desired plot values are reached. However, if adverse responses occur the operator must continue to diagnose the issue.

The diagnostic strategies vary depending on where the inefficiency or violation is occurring within the system.

- Machine schematics
- Parameter page
- Efficiency plots/loss monitors
- Alignment references (example is bullseye plot)
- Use written instructions for correcting "beam orbit"
- Expected loss/preventable loss (knowledge in the head

### 7.3.5 Task 5. Act to Prevent or Restore the Machine to Operate with Expected Efficiency Range

The fifth high-level goal is the operator acting on the system to restore the beam characteristic to a desired setting. Within this task is a series of acting, monitoring then acting again to ensure that actions taken are having the desired effect. Operators monitor using the equipment plot window and act on a parameter page by either increasing a decreasing a parameter in increments to achieve the desired result. Loss plots and other "first indications" are reviewed to ensure their actions solved the noted issue.

- Knob plots and parameter page
- Adjusting switchyard setting (directing beam to machines) to increase/decrease beam quantity
- Check experiment request (in elogs)
- Loss plots

## 7.4 Conclusion

The purpose of the operator task analysis is to document the typical operator workflow to provide a basis of inputs for functional and design requirements. It is the intention of the human factors team to conduct task analyses for additional accelerator personnel roles.

# 8.0 APPENDIX C – INDEX OF REQUIREMENTS

| Section | Sub-Section | No. | Requirement |
|---|---|---|---|
| **Global Requirements** | | | |
| Information Architecture and Navigation | Information Architecture | [HFR-1] | A two-level hierarchy should be used for system display organization. |
| Information Architecture and Navigation | Information Architecture | [HFR-2] | Level 1 information should provide general accelerator information that supports situation awareness of equipment status and beam health. |
| Information Architecture and Navigation | Information Architecture | [HFR-3] | Level 1 information should be presented on the 'Comfort Display' to rapidly communicate degrading, abnormal, or emergency conditions from "at-a-glance." |
| Information Architecture and Navigation | Information Architecture | [HFR-4] | Level 2 information should support the operator in performing their tasks to accomplish the functions assigned to them. |
| Information Architecture and Navigation | Navigation | [HFR-5] | The system should provide an index page that is organized by 1) function and 2) frequency of use for alternative navigation. |
| Information Architecture and Navigation | Navigation | [HFR-6] | System navigational options should be visible on all pages. |
| Information Architecture and Navigation | Navigation | [HFR-7] | The system should provide visual cues to inform user of where they are in the system. |
| Information Architecture and Navigation | Navigation | [HFR-8] | The system should clearly differentiate navigational elements from each other but visually group similar navigational elements together. |
| Information Architecture and Navigation | Windows, Pop-ups, and Faceplates | [HFR-9] | The content provided on all windows, pop-ups, and faceplates of the system should be located in a consistent location. |
| Information Architecture and Navigation | Windows, Pop-ups, | [HFR-10] | Pop-ups and faceplates should be presented in a location that does not obscure information related to operating equipment. |

| Section | Sub-Section | No. | Requirement |
|---|---|---|---|
| Architecture and Navigation | and Faceplates | | |
| Information Architecture and Navigation | Menus and Search | [HFR-11] | The primary display page should provide a search capability. |
| Information Architecture and Navigation | Menus and Search | [HFR-12] | The system should provide a drop-down menu across all display pages. |
| Information Architecture and Navigation | Menus and Search | [HFR-13] | The drop-down menu should be located in the top-left corner of the header across all display pages. |
| Information Architecture and Navigation | Menus and Search | [HFR-14] | A drop-down menu should contain category labels from which the drop-down menus is accessed. |
| Display Formatting | Display Layout | [HFR-15] | All display pages should contain a header with a unique title at the top of the page. |
| Display Formatting | Display Layout | [HFR-16] | All display pages should provide a navigation menu at the top left within the header. |
| Display Formatting | Display Layout | [HFR-17] | All display pages should provide a selectable breadcrumb. |
| Display Formatting | Display Layout | [HFR-18] | All display pages should provide primary canvas area that is consistently sized to support the user's primary task. |
| Display Formatting | Display Layout | [HFR-19] | All display pages should provide a unique display number that can be referenced through search entry. |
| Display Formatting | Display Layout | [HFR-20] | The system should provide the capability of saving custom display configurations (e.g., multiple graphs). |
| Display Formatting | Display Layout | [HFR-21] | Custom display configurations should be tied to a user's login. |
| Display Formatting | Display Layout | [HFR-22] | The means of saving a display configuration should be explicitly visible to the user. |
| Display Formatting | Color | [HFR-23] | Color should be used appropriately to indicate meaning. |
| Display Formatting | Color | [HFR-24] | Implicit color associations of an accelerator complex should inform overall color usage in the digital control system interface. Secondary color associations that are common in everyday life and cultural applications should also be considered. |
| Display Formatting | Color | [HFR-25] | A dull screen color scheme should be adopted to reduce display color saturation and saliency. |
| Display Formatting | Color | [HFR-26] | Saturated colors should be reserved to indicate special meaning. |
| Display Formatting | Color | [HFR-27] | Highest priority information (e.g., text or other display elements) must be tested for color blind safety. |
| Display Formatting | Color | [HFR-28] | Color should be used appropriately to indicate meaning. |

| Section | Sub-Section | No. | Requirement |
|---|---|---|---|
| | | | Details: Each color included in an accelerator interface should only have one meaning consistently throughout the system. Multiple meanings used for a single-color can cause confusion and performance issues. When multiple colors are used in a single interface, users must remember each of the different meanings in addition to their work tasks. The reliance on a user's memory is even further complicated when a single color is used to convey multiple meanings. To improve the comprehensibility of color-coding associations in accelerator interfaces, each color must be linked to a single meaning or purpose. *Technical Basis:* 0 (Section 6.3.1); 0 (Sections 7.2.5.1 – 7.2.5.10); 0 (Section 1.3.8) 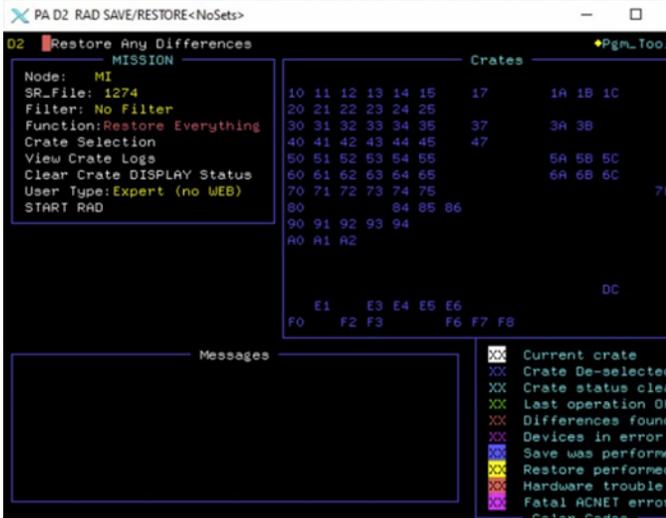 *Figure 9. Poor example of color use. One color (I.e., teal blue) is used to indicate multiple meanings (I.e., "crate status clear" and general text/headings). Each color should only have one meaning and general text colors shouldn't be reused to indicate a special status.* Once a color has been assigned a meaning, that meaning should remain consistent throughout all aspects of the interface. For example, if red is used to signal an alarm, all uses of red throughout the interface should denote the same meaning and user conclusions; whenever red appears, there's an alarm in the system. This creates consistency in the interface which helps users intuitively understand system interactions. **[HFR-76]** **Implicit color associations of an accelerator complex should inform overall color usage in the digital control system interface. Secondary** |

| Section | Sub-Section | No. | Requirement |
|---|---|---|---|
| | | | **color associations that are common in everyday life and cultural applications should also be considered.**<br><br>*Details:* Implicit color associations of an accelerator complex should inform overall color usage throughout interface design. An example of this is when Fermilab accelerator structures are painted unique colors and those colors are replicated within the control system interfaces (e.g., the New Muon Lab building interior is painted mint green and the control system interfaces have the same color as a background). A better design for this example would be to include color associated headings (instead of background) to implicitly link accelerator infrastructure to relevant control system interfaces. Technical *Basis:* 0 (Section 6.3.1); 0 (Sections 7.2.5.1 – 7.2.5.10); 0 (Section 1.3.8)<br><br>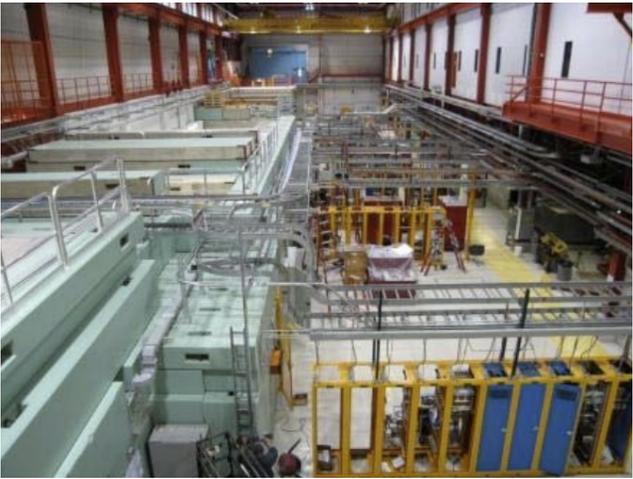<br><br>*Figure 10. New Muon Lab, interior paint color mint green.* |

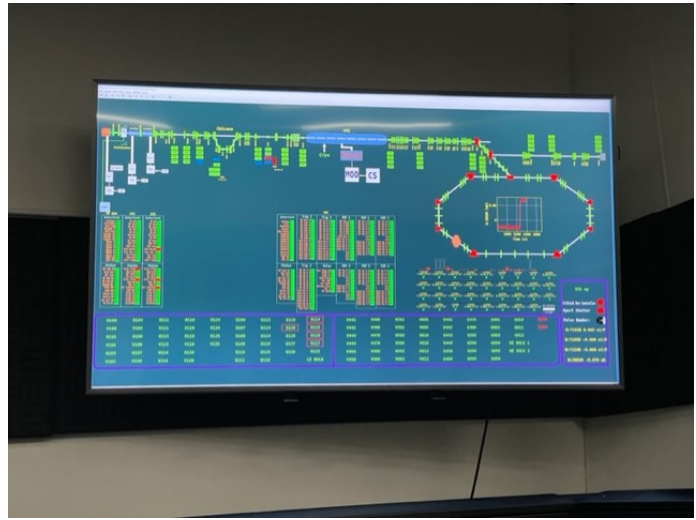

*Figure 11. New Muon Lab overview interface, background color mint green.*

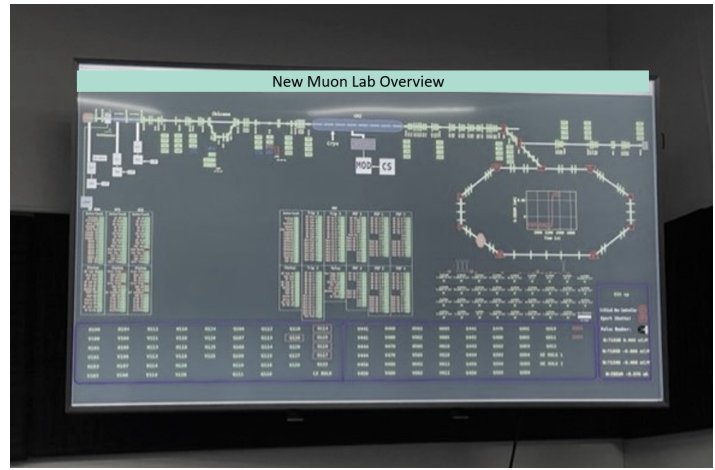

*Figure 12. Alternative design for implicit color association.*

[HFR-77] **A dull screen color scheme should be adopted to reduce display color saturation and saliency.**

*Details:* The "Dull Screen" approach is an interface design concept based on the theory that all normal behavior should appear "dull" so that abnormal behavior detected by the system can be highlighted or made salient through the use of color. 0. This strategy helps users rapidly detect events that require their

| Section | Sub-Section | No. | Requirement |
|---|---|---|---|
| | | | detailed attention. This concept can also help reduce the amount of saturated colors included in an interface which improves a user's ability to differentiate between levels of information and focus on what is most important. Technical *Basis:* 0 (Section 6.3.1); 0 (Sections 7.2.5.1 – 7.2.5.10); 0 (Section 1.3.8) 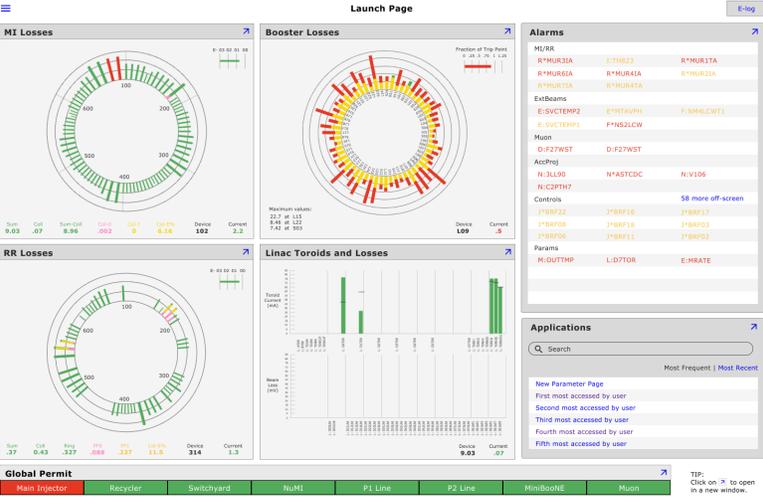 *Figure 13. Dull screen prototype (launch page).* |
| | | [HFR-78] | **Saturated colors should be reserved to indicate special meaning.** *Details:* Each color included in an interface competes with other colors and other display elements (e.g., text) for the user's attention. In line with the dull screen concept, highly saturated colors should be reserved for special or critical elements of a display. Special or critical elements of a display are those that must effectively draw a user's attention. In an accelerator digital control system, this can include multiple elements such as live data that must be continuously monitored or alarm events that require immediate attention. However, when too many elements on a display use color to convey special meaning, all information included in the display is disguised as equally important. In other words, if everything appears special, nothing appears special. Reserving saturated colors to indicate special meaning for certain display elements improves overall user performance and strengthens the comprehensibility of interface color- |

| Section | Sub-Section | No. | Requirement |
|---|---|---|---|
| | | | coding associations. *Technical Basis:* 0 (Section 6.3.1); 0 (Sections 7.2.5.1 – 7.2.5.10); 0 (Section 1.3.8) |
| | | [HFR-79] | **Highest priority information (e.g., text or other display elements) must be tested for color blind safety.** *Details:* Color blindness is the decreased ability to see color or differences in color within afflicted individuals. Color blindness affects about eight percent of males (approximately 10.5 million) and less than one percent of females. 0. There are two major types of color blindness: those who have difficulty between red and green, and those who have difficulty distinguishing between blue and yellow. A challenge in designing to accommodate color blindness is trying to accommodate the unknown. Not only do color-blindness types vary, but the level of color discernment ability varies between individuals as well. Although estimates of afflicted individuals are known, it is difficult to ascertain which color-blindness type potential users might have as well as what their level of color discernment is. Research models have been established to predict or calculate how colors are perceived by color-blind people, but they are not completely accurate. In other words, it isn't possible to predict future end user color blindness type and variability with 100% accuracy. Fortunately, there are some baseline recommendations that regardless of color blindness type and variability, will help accommodate color blindness. |

| Section | Sub-Section | No. | Requirement |
|---|---|---|---|
| | | | *Figure 14. Color blindness types and color combinations to avoid.*<br><br>The more important the interface content is, the more essential it is to make it color blind safe. Color blind safety is a concept that encourages certain types of color use to accommodate color blindness. The colors most detectable by anyone with color blindness are black and white (e.g., black text/elements on a white background and vice versa). This is because these colors have the highest contrast ratio compared to all other colors and are easily discernable from each other. If additional colors must be used, interface content areas should be monochromatic with the interface element color and background color at the opposite ends of the color saturation poles. Refer to the color palettes presented in Table 1 and Table 3. *Technical Basis:* 0 (Section 6.3.1); 0 (Sections 7.2.5.1 – 7.2.5.10); 0 (Section 1.3.8)<br><br>The system should apply the project defined color palette consistently across control system display pages. |
| Display Formatting | Text: Titles, Labels, Abbreviations, and Acronyms | [HFR-29] | All alphanumeric text (static and dynamic) should be no less than 9-point font (or 16 minutes of arc) for adequate legibility. |
| Display Formatting | Text: Titles, Labels, Abbreviations, and Acronyms | [HFR-30] | All alphanumeric text should be Verdana. |

| Section | Sub-Section | No. | Requirement |
|---|---|---|---|
| Display Formatting | Text: Titles, Labels, Abbreviations, and Acronyms | [HFR-31] | All alphanumeric text variations should be consistent throughout all interfaces. |
| Display Formatting | Text: Titles, Labels, Abbreviations, and Acronyms | [HFR-32] | The system should use a label convention that is intrinsically meaningful to the users. |
| Display Formatting | Text: Titles, Labels, Abbreviations, and Acronyms | [HFR-33] | The labeling convention should be applied consistently across the system. |
| Display Formatting | Text: Titles, Labels, Abbreviations, and Acronyms | [HFR-34] | All labels to be read should be oriented horizontally on display pages. |
| Display Formatting | Text: Titles, Labels, Abbreviations, and Acronyms | [HFR-35] | The system should visually differentiate between labels that are clickable and labels that are information only. |
| Display Formatting | Dynamic Display Element Formatting | [HFR-36] | Clearly distinguish contextual information from live plot data. |
| Display Formatting | Dynamic Display Element Formatting | [HFR-37] | Present only necessary data on a plot to improve user time to complete task or understand system status. |
| Display Formatting | Dynamic Display Element Formatting | [HFR-38][HFR-36] | Clearly distinguish contextual information from live plot data. Details: Presenting operational context within plots (see figure 22) supports rapid recognition of system status. As adding operational context increases the information density of a plot, the design should make distinguishing the operational context from the live data easy to recognize. ACNET often displays new data over the old to help show change over time. However, there is little by way of design to distinguish the entry time of each new reading. Distinguishing the age of the reading in the plot may lead to some emergent features and understanding by users. Technical *Basis: 0* |

| Section | Sub-Section | No. | Requirement |
|---|---|---|---|
| | | | 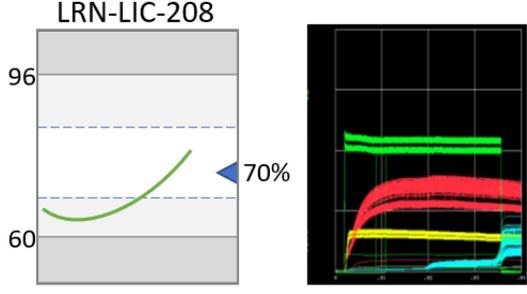 *Figure 22. The left figure shows live data (green, salient) against its operational context (light gray, neutral). The right figure shows data collected over many retrievals, but not indication of the timeline of retrieval.*<br><br>**[HFR-80]** **Present only necessary data on a plot to improve user time to complete task or understand system status.**<br>*Details:* Time to complete a task increases with task complexity. The more complex a task the more important it is to determine exactly what information is required. Tendencies to provide extra information can only increase the user need to process and interpret what is important and necessary from the extra information available. Developing a plot to fit a task will improve user's ability to perform the task. However, this pertains to novice users performing regular tasks. The capability to craft plots based on an emergent need should be available. Technical *Basis: 0*<br><br>Graphs and charts should include labels for its title, axes, parameters, and engineering units. |
| Display Formatting | Dynamic Display Element Formatting | [HFR-39] | Graphs and charts should include a digital readout of the parameter(s) being represented when precise reading is required of the user. |
| Display Formatting | Dynamic Display Element Formatting | [HFR-40] | Where multiple are presented on a single graph or chart, each parameter should be coded through the use of color or line type for differentiation. |
| Display Formatting | Dynamic Display | [HFR-41] | Where multiple graphs and charts are compared on a single display page, their x-axes and y-axes should consist of the same scale. |

| Section | Sub-Section | No. | Requirement |
|---|---|---|---|
| | Element Formatting | | |
| Display Formatting | Dynamic Display Element Formatting | [HFR-42] | All components, line points, and termination points presented on a mimic display should be labeled. |
| Display Formatting | Dynamic Display Element Formatting | [HFR-43] | Flow direction should be coded with distinctive arrowheads. |
| Controls and Interaction | User Input and Controls | [HFR-44] | The primary input device mode should be cursor-based (i.e., a mouse). |
| Controls and Interaction | User Input and Controls | [HFR-45] | The secondary input device mode should be keyboard-based (i.e., use of shortcuts for navigation). |
| Controls and Interaction | User Input and Controls | [HFR-46] | The system should provide indication of all display elements that include control functionality. |
| Controls and Interaction | User Input and Controls | [HFR-47] | All control options for a specific soft controller (i.e., faceplate) should be made accessible by a single click. |
| Controls and Interaction | User Input and Controls | [HFR-48] | All frequency performed control actions should be accessible from a soft control faceplate without any additional administrative action. |
| Controls and Interaction | User Input and Controls | [HFR-49] | Soft control options should be suitable for characteristics of the task performed, by using Table 5. |
| Controls and Interaction | User Input and Controls | [HFR-50] | Soft controls should be visually distinguishable from administrative buttons like navigation buttons. |
| Controls and Interaction | User Input and Controls | [HFR-51] | Data entry should be accompanied by a verification step. |
| Controls and Interaction | User Input and Controls | [HFR-52] | The system should provide the capability of creating a control script for complex sequence of actions. |
| Controls and Interaction | User Input and Controls | [HFR-53] | The system should enable the user to correct for erroneous entries during data entry. |
| Controls and Interaction | User Input and Controls | [HFR-54] | The system should provide confirmations for control actions that are safety important or have potential to disrupt normal operation. |
| Controls and Interaction | User Input and Controls | [HFR-55] | The system should provide a means of restoring user defined settings in the event of a system failure. |
| Controls and Interaction | User Input and Controls | [HFR-56] | The system should prohibit multiple users from controlling the same equipment. |
| Controls and Interaction | System Interaction | [HFR-57] | Visual feedback should be provided across all user interactions with the system. |

HUMAN-SYSTEM INTERFACE STYLE GUIDE FOR ACORN DIGITAL CONTROL SYSTEM

| Section | Sub-Section | No. | Requirement |
|---|---|---|---|
| | and Feedback | | |
| Controls and Interaction | System Interaction and Feedback | [HFR-58] | Visual feedback should be applied consistently across the control system. |
| Controls and Interaction | System Interaction and Feedback | [HFR-59] | System latency should be 0.2 seconds or less for real-time responses. |
| Controls and Interaction | System Interaction and Feedback | [HFR-60] | The system should indicate that a user's input is processing for system response times greater than 1 second. |
| Controls and Interaction | System Interaction and Feedback | [HFR-61] | Blinking/flashing should be used only for alerting the user to events that require immediate attention. |
| Controls and Interaction | System Interaction and Feedback | [HFR-62] | No more than two blink/flash rates should be used. |
| Controls and Interaction | System Interaction and Feedback | [HFR-63] | Alarms should be used only for off-normal conditions that require timely action by the user. |
| Controls and Interaction | System Interaction and Feedback | [HFR-64] | The system should provide a means to suppress alarms based on expected conditions from an experiment. |
| Controls and Interaction | System Interaction and Feedback | [HFR-65] | The system should provide the user with notifications of any conditions that may impact the accelerator's performance. |
| Controls and Interaction | System Interaction and Feedback | [HFR-66] | The system should provide an indication that the display is reading data from the control system (i.e., system heartbeat). |
| **Function-Based Requirements** | | | |
| Accelerator Displays | Human Factors Functional Requirements for Monitoring, Detection, and Selection | [HFR-67] | The system should provide contextual status of 1) alarms 2) relative losses of critical machines 3) and access to an event summary report over the past shift to support shift briefings. |
| Accelerator Displays | Human Factors Functional Requirement | [HFR-68] | The system should provide status of available accelerator parameters deemed important for beam health monitoring across the accelerator complex. |

| Section | Sub-Section | No. | Requirement |
|---|---|---|---|
| | s for Monitoring, Detection, and Selection | | |
| Accelerator Displays | Human Factors Functional Requirements for Monitoring, Detection, and Selection | [HFR-69] | The system should support the capability for a user to select available accelerator parameters from the control system that are deemed important for monitoring but not part of the standards MDS display loadout. |
| Accelerator Displays | Human Factors Functional Requirements for Monitoring, Detection, and Selection | [HFR-70] | The system should provide accessibility to the electronic logs (Elogs) for detailed descriptions of previous malfunctions, troubleshooting, maintenance, and issue resolutions. |
| Accelerator Displays | Human Factors Functional Requirements for Monitoring, Detection, and Selection | [HFR-71] | The system should enable the user to monitor accelerator status and overall beam quality without any administrative task burden. |
| Accelerator Displays | Human Factors Functional Requirements for Monitoring, Detection, and Selection | [HFR-72] | The system should notify users of machine/ equipment conditions that are indicative of negatively impacting beam characteristics requiring adjustment. |
| Accelerator Displays | Human Factors Functional Requirements for Monitoring, Detection, and Selection | [HFR-73] | The system should allow users to access detailed information of any abnormal machine/ equipment condition that requires adjustment. |
| Accelerator Displays | Human Factors Functional | [HFR-74] | Displays should indicate when data presented is live or no longer up-to-date |

| Section | Sub-Section | No. | Requirement |
|---|---|---|---|
| | Requirements for Monitoring, Detection, and Selection | | |
| Accelerator Displays | Human Factors Functional Requirements for Monitoring, Detection, and Selection | [HFR-75] | The MDS page should support both proactive and reactive responses to accelerator statuses that may or currently require operator attention |
| **Design Requirement Specifications** for Monitoring, Detection, and Selection | | | |
| Launch Page | Design Requirement Specifications | [DRS-1] | A single landing page should be designed for all users performing high-level monitoring of the accelerator complex. |
| Launch Page | Design Requirement Specifications | [DRS-2] | Display a targeted set of dynamic, high-level information (such as plots with one or multiple parameters) for monitoring and detection purposes. |
| Launch Page | Design Requirement Specifications | [DRS-3] | Display relative losses of key accelerator systems based on expected system operating parameters established at beginning of shift or after key activities. |
| Launch Page | Design Requirement Specifications | [DRS-4] | Display a system application list determined by user usage to efficiently navigate through the system for quick resolution and task completion. |
| Launch Page | Design Requirement Specifications | [DRS-5] | Display priority alarms affecting beam to alert for operator response. |
| Launch Page | Design Requirement Specifications | [DRS-6] | Display search bar with function to locate any system application via name or identification code. |
| Accelerator Displays | Human Factors Functional Requirements for Diagnosis and Acting | N/A | This section is out of scope for this revision of the style guide. Future revisions will include display specifications for these roles. |
| Analysis and Control Displays | N/A | N/A | This section is out of scope for this revision of the style guide. Future revisions will include display specifications for these roles. |

| Section | Sub-Section | No. | Requirement |
|---|---|---|---|
| Custom Parameter Page Displays | N/A | N/A | This section is out of scope for this revision of the style guide. Future revisions will include display specifications for these roles. |
| TBD | TBD | TBD | TBD |

## 9.0 APPENDIX D – COLOR GLOSSARY

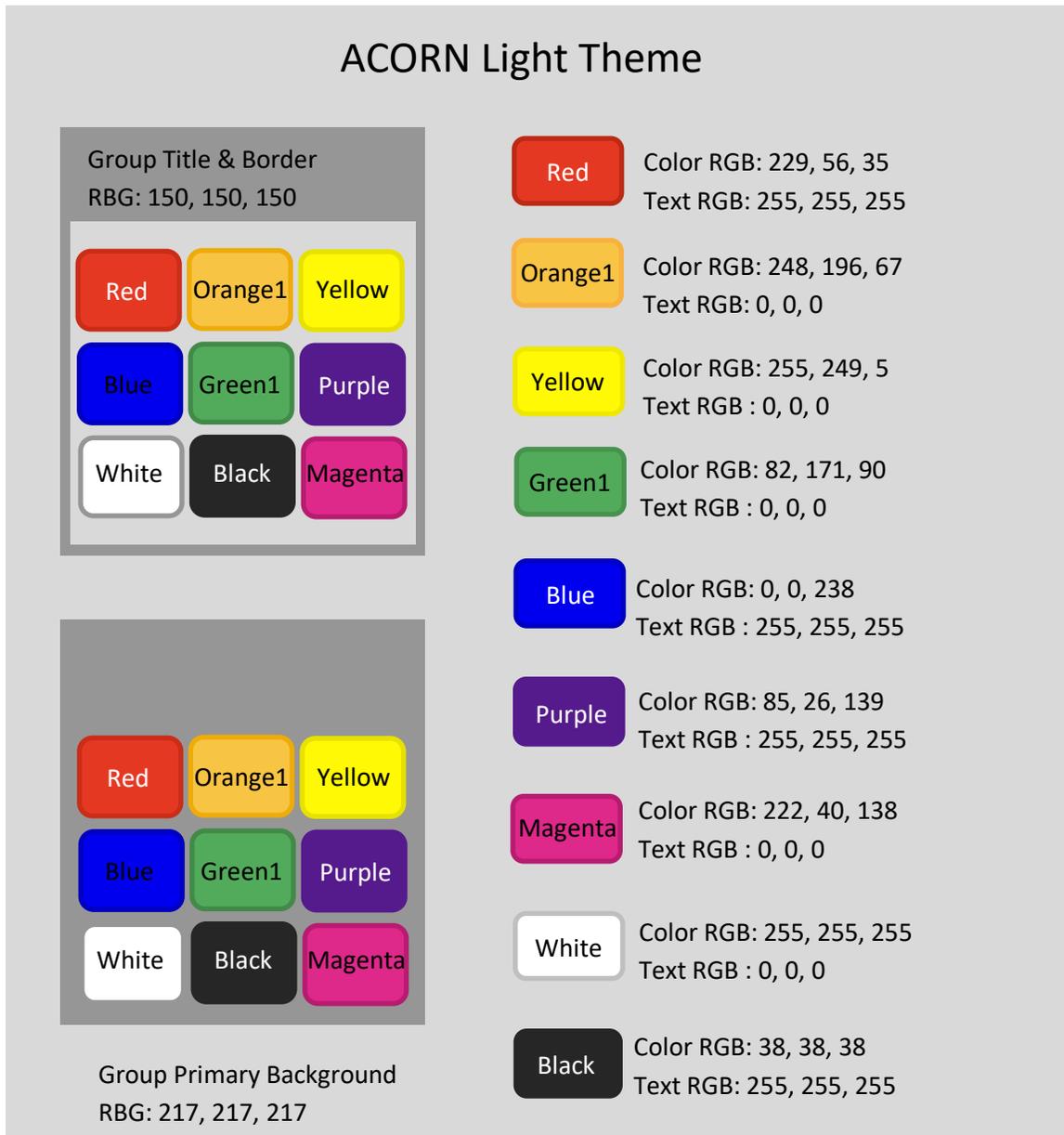

Figure 33. ACORN Light Theme Color Palette

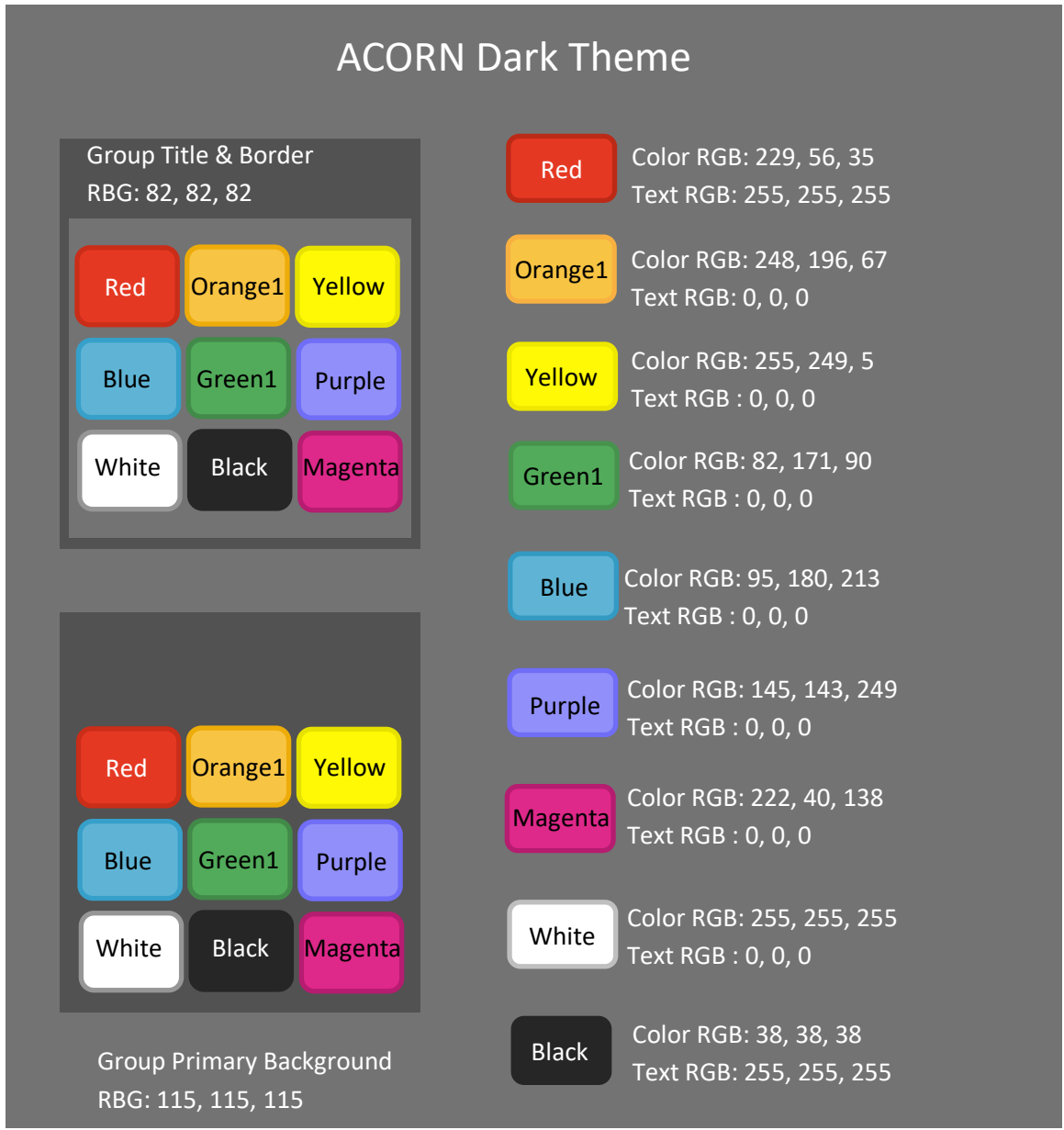

*Figure 34. ACORN Dark Theme Color Palette*

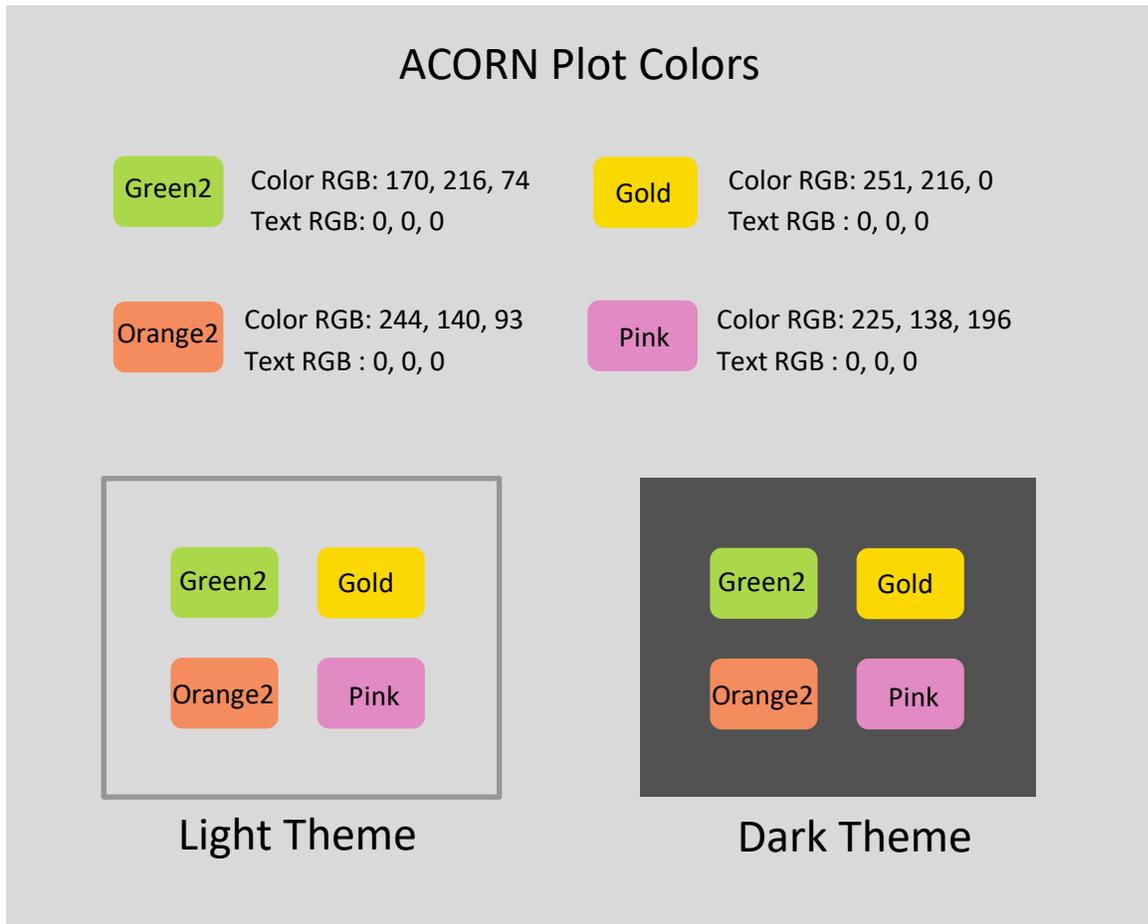

*Figure 35. ACORN Plot Color Palette*

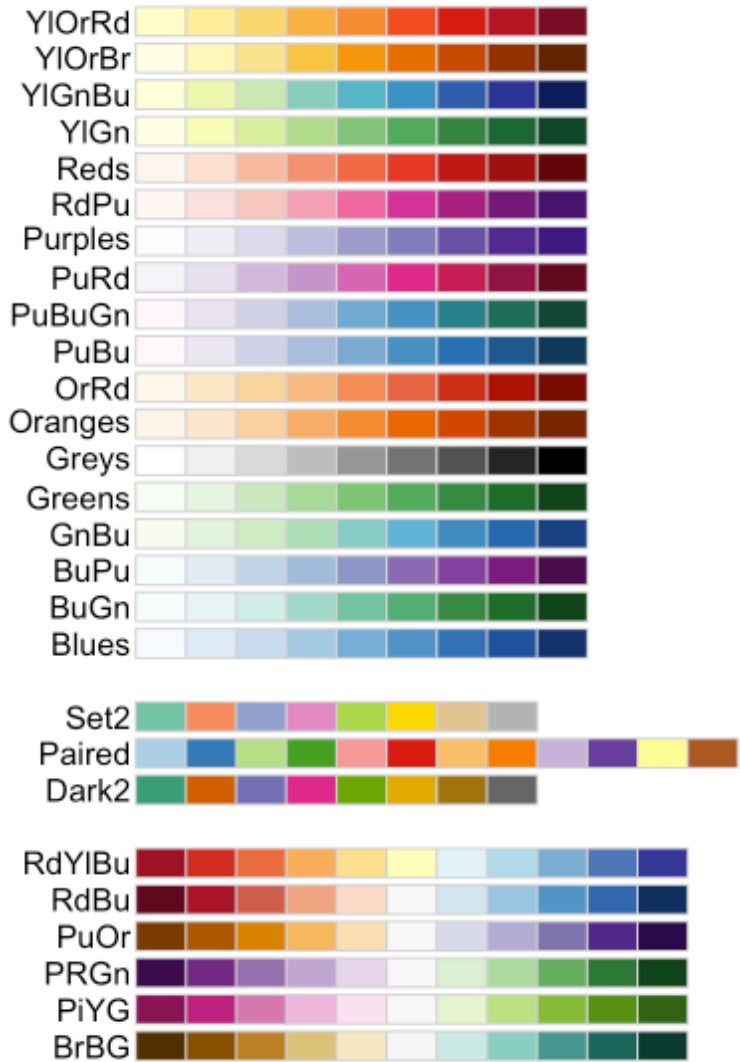

Figure 36. Colorblind-friendly RColorBrewer® Palettes for data visualization.

## 10.0 APPENDIX E – DISCUSSION OF TRADE-OFFS

Every interface has different requirements due to targeting different users, with different skill-sets, trying to accomplish different goals. This variability will ultimately lead to decisions that may sacrifice one design element in favor of another. Trade-offs of this kind are a natural part of the design process and must be considered and a direction decided upon. It can be hard to predict where trade-off decision will be required. However, there are some general principles that will likely clash during the design of an accelerator control system such as:

1. Designing for inherent and perceived usability
2. Designing for power user capability and novice user familiarity
3. Meeting user wants while designing for user needs

Inherent and perceived usability can both have implications for the user experience and their ability to perform functions. Perceived usability is the user's impression of the control system, how they perceive the interface visibility and transparency. Their feeling while using the system can also contribute to perceived usability. High perceived usability generally results in a positive experience and great initial impressions of the HSI. Inherent usability is the quantifiable capability of a system to support users in completing their goals. Inherent usability refers to the HSI's actual ability to address the gulf of evaluation and execution. Both types of usability are important aspects of a design. However, when the two conflict in a design decision it is recommended, in the case of developing an HSI for accelerator operators, to opt for inherent usability.

Both power and novice users will be interacting with essentially the same HSI design for accelerator operations. At times, it is difficult to assess which user persona takes priority in the design process. At these times, it is prudent to return to the high-level goals for the upgraded HSI design such as decreasing the time to train new control room operators. As discussed in later sections, novice users will require greater system visibility as they familiarize themselves with the accelerators and the HSI. Therefore, power user functionality can be added such that they require a little more understanding and experience with the system to use. That way as users progress from novice to power users they will gradually engage with more HSI functionality and learn how to apply it to their tasks.

Occasionally, user wants and user needs will clash. It is important to have an HSI that users want and enjoy using but ultimately it must be functional and support their ability to perform in the control room. As these trade-offs are encountered the designers must weigh the implications of the different decisions and how they impact the operator's ability to perform in the control room. If the cost to functionality is low and the benefit of improving the perceived usability is great enough, there may be justification for acquiescing to user wants.

There may be other instances where a trade-off decision is encountered. Each trade-off must be evaluated for the cost to HSI functionality and inherent usability. When these decisions are made, there must be a method for ensuring that consistency across HSIs is maintained.

# 11.0 APPENDIX F – GLOBAL USABILITY PRINCIPLES j

Many designers and developers are tempted to quickly begin designing interfaces without truly understanding what they are designing and what requirements their designs will meet. Rushing into "the how" without clearly defining "the what" causes designers to improvise. Improvising in design can lead to an endless cycle of iteration where solutions are proposed without clarity, thoughtfulness and even consensus. This not only creates inconsistency throughout designs, but also disconnection and incoherence.

On the contrary, understanding "the what" before even thinking about "the how" provides a clear path for designers towards uniformity. "The what" may somewhat change over time which is another reason this document is expected to be revised regularly. However, this document also provides both general guidance of usability principles and specific guidance of interface design specifications. The combination of these guidelines was created through consolidating findings from accelerator operator interviews, evidenced-based human factors design principles, and considerations from other accelerator control system benchmarks. Adhering to these guidelines will lay the foundations for designs that are robust and congruous.

This section introduces long established usability principles. Usability principles are design principles that when used correctly, enable intuitive and user-friendly interfaces. The stronger the usability of an interface, the better equipped users are to optimally perform. Each of the usability principles introduced should be applied to all concepts of design.

## 11.1 CONSISTENCY

Consistency in interface design is making elements uniform across a digital system. The principle of consistency creates the foundation for all design as changing the look and feel of elements from page to page dissociates the entire system. Consistency creates predictability in an interface which enables user confidence and optimizes overall performance [1]. Design consistency is also conforming to a set of standards and persistently applying them throughout all interface designs. Users should not have to wonder whether different labels, colors, or other elements mean the same thing from page to page throughout a digital system. Failing to maintain design consistency increases cognitive workload and can lead to burnout.

### 11.1.1 DESIGN CONSISTENCY ENABLES FLEXIBLE OPERATIONS

Accelerator environments are complex, and users rely on some level of flexibility to maintain efficient operations. The idea of supporting flexibility might seem contradictory to the principle of consistency, but the opposite is true: properly applied design consistency actually enables flexibility within user operations. Creating consistency in how a system looks and feels directly increases the trust and familiarity the user experiences and thus increases the potential for system mastery [1]. The more proficient a user is within a system, the greater their capability is to use the system in a way that best fits their needs (i.e., flexibly).

## 11.2 FAMILLIARITY

Jakob's law states that the time users spend interacting with technology outside of work influences how they expect the technology they interact with at work to function [2]. Therefore, designing the interaction platform for the Fermi accelerators should incorporate familiar elements found across other interactions with websites, technology, or system platforms. Doing so can improve both the inherent and apparent usability of a system. The inherent being the objective usability of a design and the apparent being the perceived usability of a design. Employing familiar concepts, icons, layouts, imagery, and tools can be especially helpful for novice users working with the system. The way the system acts and the user's expectation of system behavior will align better and earlier in the training process. This is also referred to as designing to match the user's mental model.

## 11.3 USE MENTAL MODELS

Understanding and leveraging mental models can be a powerful tool for engendering familiarity. Mental Models are constructs users have developed in their mind that represent their understanding of how a system functions. Interfaces and interactions that match popular mental models can accelerate user's familiarity with a system. Novice users working with a familiar interface can rapidly learn how to use the system. However, an HSI can also be designed to inform or reinforce a particular mental model. This technique can help novice users develop their mental model of how the accelerators work. Most people have not operated an accelerator and therefore do not have a mental model of how one must be operated. This can be challenging to create a familiar design. In this specialty case, it is prudent to design the system interface and the training material in the same way such that a *familiar* mental model can be developed and leveraged during operation.

## 11.4 SIMPLICITY

Designing simple interfaces can be thought of as the practice of maximizing meaning and information while minimizing the amount of "ink" on the page [3]. An effective simple design means the functionality required to perform a task is included, and how to perform that task is clear. Furthermore, a simple display is one that is not susceptible to misinterpretation. Simplicity does not refer to removing features or capabilities for the sake of simplicity. The number of tasks and amount of information required to run an accelerator may call for full screens and interfaces that make many capabilities always present and available. Natural mappings to interactions can improve the simplicity of an interface as well. Grouping related parameters and designing proximity-compatibility relationships between controls and parameters can improve simplicity.

Also, when designing for simplicity, consider the mix of expert and novice users. The interface should focus on supporting novice user tasks directly on the interface in a clear and obvious way. Tasks suited or geared for expert users may require an additional step to access or slightly more nuanced interaction. As users grow from novice to expert users these features will become more explored as the transition occurs offering a natural development from novice to expert. Designing for gradual engagement of this nature requires specific intention as to how the interface may be designed.

## 11.5 ABSTRACT AND AGGREGATE DATA

Abstracting data to higher-level meaning offers a means to support users more directly by creating immediately meaningful information that requires less cognitive processing than presenting raw data. Abstraction can be applied to labels and telemetry but using natural language or providing context respectively. Both of which reduce interpretation and provide more direct input to the user.

Aggregating data is a simple method to take related inputs and combine them into a single indication. Common applications of aggregating data are interlock decision trees. The decision tree can be consolidated to a single indication of 'met' or 'unmet' clearing space on the screen for other relevant and meaningful elements. If the desired status is not displayed the option to investigate should be present. In any instance of aggregating data, the ability to individually investigate the aggregated elements should be present especially when expert users are involved.

## 11.6 TRANSPARENCY

The system interface should feel like a direct link to the system. Transparency typically refers to the opacity of an object or the ability to see through an object. When applying transparency to interface design it takes on a slightly different meaning. Transparency is a design concept to overcome the "gulf of execution" [4]. The gulf of execution refers to the challenge of performing an action to accomplish a task. A transparent interface links the user directly to their tasks and the feedback the system provides when performing their tasks. It removes anything that can distract or add extra steps for a user when trying to perform an action. A transparent display is also referred to as having a high correspondence to the system it is designed for. That is the mapping of interface to system domain. If the interface and system have a high correspondence the transparency of the interface is increased. Correspondence is an interaction between the system and the interface. Therefore, it is equally as important that after an action is made the system provides immediate feedback of having received the action.

## 11.7 VISIBILITY

System visibility is the capability of an interface to communicate system status accurately and quickly to the user. A highly visible system overcomes the "gulf of evaluation" or the user ability to understand the system state to decide what action is necessary [4]. This concept is a parallel to system transparency. The two together are comprehensive in 1. Evaluate the system and 2. Take proper action. Then the cycle repeats. As highly transparent systems have good correspondence to the system, interfaces with high system visibility have good coherence. Coherence refers to how the graphical elements, lay out, and design of the system come together to clearly communicate system status to the user quickly and accurately. One tactic to improve coherence and achieve visibility is by leveraging human perceptual processing.

## 11.8 LEVERAGE PERCEPTUAL PROCESSING

Perceptual processing is a term cognitive psychologists use when referring to how humans sense their environment and then add meaning to it based on current awareness and past experience. When designing an interface this refers to adding inherent meaning to the interface that reduces the need for the user to perform this cognitive processing stage. The symbology, iconography, plots, text, and other design elements should be coherent with the system in operation. For expert users, this strategy creates greater opportunities to see larger patterns or associate deeper meaning to what they are seeing. For novice users, this reduces the need for training as the meaning and feedback provided in the interface acts as a training device.

## 11.9 EASE OF USE

Ease of Use is the coalescence of all other usability principles brought together. It is the measure of how well a system supports the operator through monitoring, evaluation, and task execution. It can also refer to user satisfaction and enjoyment while using interacting with the system. Ease-of-use can be a moving target however, especially when users with different levels of experience must use the same system. Systems designed for novices typically do not overwhelm the user with available functionality visible on the screen opting to select only the basic functionality that a novice may need. This design may reduce the ease-of-use for an expert or power user. So, trade-offs must be assessed when designing for ease of use.  Ease-of-use is a measure of how users perform their goals with as little burden or intermediary steps between evaluation and execution.

## 11.10    Global Usability Principles References

1. Krause, R. (2021). Maintain Consistency and Adhere to Standards: Usability Heuristic #4. Nielsen Norman Group
2. Nielsen, J. (2000, July 22). End of Web Design. Nielsen Norman Group. URL:https://www.nngroup.com/articles/end-of-web-design/ (Nielsen, J., 2000).